\documentclass[acmsmall]{acmart}

\setcopyright{rightsretained}
\acmJournal{PACMHCI}
\acmYear{2024} \acmVolume{8} \acmNumber{ETRA} \acmArticle{236} \acmMonth{5}\acmDOI{10.1145/3655610}

\makeatletter
\gdef\@copyrightpermission{
  \begin{minipage}{0.2\columnwidth}
    \href{https://creativecommons.org/licenses/by-nc-sa/4.0/}{\includegraphics[width=0.90\textwidth]{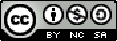}}
  \end{minipage}\hfill
  \begin{minipage}{0.8\columnwidth}
    \href{https://creativecommons.org/licenses/by-nc-sa/4.0/}{This work is licensed under a Creative Commons Attribution-NonCommercial-ShareAlike 4.0 International License.}
  \end{minipage}
  \vspace{5pt}
}

\usepackage{multirow} 
\usepackage{enumitem}
\usepackage{graphicx}
\usepackage{subcaption}
\usepackage{geometry}
\newcommand{\dataset}{\textsf{HCEye}}
\newcommand\changed[1]{#1}

\begin{document}

\title[Shifting Focus with HCEye]{Shifting Focus with HCEye: Exploring the Dynamics of Visual Highlighting and Cognitive Load on User Attention and Saliency Prediction}

\author{Anwesha Das}
\email{adas@cs.uni-saarland.de}
\orcid{0009-0006-8308-9961}
\authornote{These authors contributed equally to this work. This remark is missing in the version on the ACM DL.}
\affiliation{%
 \institution{Saarland University, Saarland Informatics Campus}
 \city{Saarbrücken}
 \state{Saarland}
 \country{Germany}
 \postcode{66123}
}
\author{Zekun Wu}
\email{wuzekun@cs.uni-saarland.de}
\orcid{0000-0002-5233-2352}
\authornotemark[1]
\affiliation{%
 \institution{Saarland University, Saarland Informatics Campus}
 \city{Saarbrücken}
 \state{Saarland}
 \country{Germany}
 \postcode{66123}
}
\author{Iza Škrjanec}
\email{skrjanec@lst.uni-saarland.de}
\orcid{0009-0001-7044-8957}
\affiliation{%
 \institution{Saarland University, Language Science and Technology}
 \city{Saarbrücken}
 \state{Saarland}
 \country{Germany}
 \postcode{66123}
}
\author{Anna Maria Feit}
\email{feit@cs.uni-saarland.de}
\orcid{0000-0003-4168-6099}
\affiliation{%
 \institution{Saarland University, Saarland Informatics Campus}
 \city{Saarbrücken}
 \state{Saarland}
 \country{Germany}
 \postcode{66123}
}


\begin{abstract}
Visual highlighting can guide user attention in complex interfaces. However, its effectiveness under limited attentional capacities is underexplored. This paper examines the joint impact of visual highlighting (permanent and dynamic) and dual-task-induced cognitive load on gaze behaviour. Our analysis, using eye-movement data from 27 participants viewing 150 unique webpages reveals that while participants' ability to attend to UI elements decreases with increasing cognitive load, dynamic adaptations (i.e., highlighting) remain attention-grabbing. The presence of these factors significantly alters what people attend to and thus what is salient. Accordingly, we show that state-of-the-art saliency models increase their performance when accounting for different cognitive loads. Our empirical insights, along with our openly available dataset, enhance our understanding of attentional processes in UIs under varying cognitive (and perceptual) loads and open the door for new models that can predict user attention while multitasking.
\end{abstract}




\maketitle

\section{Introduction}
% ANNA I added a first structure for the intro here, feel free to contribute text that is not entirely AI generated please... 

%General Introduction to the Topic
Visual highlighting has been established as a tool to guide users' attention in complex user interfaces (UI) by changing visual attributes such as colour, hue, or movement (e.g. ~\cite{feit20,Findlater2009,fisher89,ostkamp2012}). Previous research has focused on designing visual cues through empirical studies~\cite{Findlater2009,mairena2022emphasis,philipsen94,robinson11,strobelt16}.
The recent advancement of deep-learning-based saliency models, which can predict fixations of users based on images of a UI~\cite{jiang2023ueyes, leiva2020understanding}, show promise to automate the design and placement of visual cues~\cite{muller22} in adaptive interfaces that direct a user's attention to information relevant to their task or context~\cite{feit20,Findlater2009}.
However, it is unclear whether such image-based saliency models can predict shifts in user attention due to dynamic changes in the UI.
Moreover, we see a gap in understanding the effectiveness of visual highlighting in relation to the user's cognitive state. Given the prevalence of multitasking, the cognitive load induced by other tasks may influence the user's gaze behaviour ~\cite{lavie2005distracted} and thus the efficacy of visual highlighting.

%Goal of the paper 
In this paper, we \changed{explore} the joint impact of visual highlighting and cognitive load on users' gaze behaviour when viewing webpages. Our goal is twofold: (1) to empirically understand how the dynamics of highlighting and cognitive load affect users' viewing behaviour of interfaces and the effectiveness of visual cues, and (2) to evaluate whether state-of-the-art saliency models can predict users viewing behaviour of UIs under cognitive load and in the presence of visual highlighting. 

To this end, we collect a new dataset, \dataset{}, containing gaze data from 27 people looking at 150 different website screenshots, displayed under a combination of different conditions: with or without highlighting, which is shown permanently or appears dynamically after 3s; without any secondary task or while performing a non-visual secondary task (counting out loud) that induced either high or low cognitive load. 
Given the complex study design, we conduct a thorough statistical analysis of the collected data using generalized linear mixed models (GLMMs), examining how visual highlighting and cognitive load jointly impact user attention, \changed{disentangling the effect due to confounding factors, such as individual differences and UI design complexities, thereby addressing a key gap in existing literature.}

%Key Findings
% Provide a high-level overview of the key findings
% Mention the influence of these factors on what is salient and how it relates to state-of-the-art saliency models.

Our study reveals, the presence of visual highlighting on a previously unnoticed region of a complex webpage not only increased its noticeability but also significantly influenced participants' gaze behaviour on the overall webpage. \changed{Participants engaged longer with the highlighted region than any other, even those that were `naturally' salient like captivating images or titles and thus explored less of the webpage.} Their ability to explore a wider number of regions was also impacted when under cognitive load, due to longer information processing times, also reducing the effectiveness of visual highlights. Yet, dynamic highlighting effectively countered this, sustaining the noticeability of highlighted regions.
%whilst allowing for quicker
% These findings imply the need for saliency models to consider dynamic factors, such as visual highlighting or other visual adpatations. Adopting in predicting user attention

%What do we do about saliency models
Our empirical findings motivate the need to develop computational models of visual saliency that are tuned towards the cognitive state of users and the dynamics of the interface. 
Therefore, we implement a state-of-the-art saliency model, SimpleNet\cite{reddytidying} and show that by adjusting the input structure of the training data and tuning on the \dataset{} dataset from the respective experiment condition, the model significantly increases the predictive performance of users' fixation in different highlight and cognitive loads. 

We hope that our dataset and findings will encourage future research to shift their focus 
%from exploring saliency of different design types (apps, webpages, etc.) 
to developing predictive models that can account for dynamic content in UIs and the impact of multitasking which might induce cognitive load. These could power adaptive UIs that support people in efficiently distributing their attention through dynamic visual changes. 
%Contribution summary
In summary, this paper contributes:
\begin{itemize}[topsep=0pt]
    \item \dataset{}, a novel gaze dataset of 150 unique webpages viewed under a variety of conditions covering different highlighting techniques and cognitive loads, overall 1350 stimuli. 
    \item New empirical insights on the effects of visual highlighting and cognitive load on user attention, disentangled from other confounding factors through statistical modelling with GLMMs. 
    \item Saliency models that predict visual attention on webpages with dynamic content viewed under different cognitive loads. 
\end{itemize}

\section{Background}

\subsection{Visual Cues in User Interfaces}

Human attentional mechanisms prioritize specific elements from incoming visual stimuli for deeper analysis, \changed{by either directing attention to salient regions (bottom-up attention) or by choosing areas that are of interest to the viewer due to prior knowledge about a stimulus or related task (top-down attention)}~\cite{treisman1980feature,rosenholtz2012rethinking}. In bottom-up attention, the peripheral vision plays a crucial role in guiding attention through basic cues like movement, luminance changes, and edge identification~\cite{strasburger2011peripheral}. \changed{In contrast, top-down attention stems from a person's goals or prior experience which biases attention towards regions aligning with the person's expectations. In practice, human attention on a stimulus is the result of both mechanisms.} While most GUI layouts are already optimized for user experience, reflecting top-down considerations, such as target search, more recent efforts in UI design aim to guide attention to \changed{incoming} relevant information using visual cues that attract bottom-up attention \cite{de2009towards, feit20, Findlater2009}, like highlighting Areas of Interest (AOIs) using colours (hue, intensity, and saturation). These proved more effective in manipulating the saliency of UI elements than other cues, such as changing shape, size, or position~\cite{robinson2011highlighting, mairena2022emphasis}, whilst also reducing reaction times for target identification in safety critical interfaces~\cite{nicosia2021design, schriver2017expertise}. We employ colour highlights \cite{strobelt2015guidelines} in our study and analyze how their effectiveness is modulated by the dynamics of the highlight and the cognitive load of the viewer.

\subsection{Influence of Cognitive Load on Visual Attention}
Our eyes, intricately linked to the nervous system, provide insights into cognitive processes and internal states \cite{kramer2020physiological}. Whilst engaging in mentally demanding tasks like complex math problems, observable visual patterns emerge — increased blinking, frequent downward gazes, and reduced fixations indicating heightened cognitive load \cite{chen2023high, lavie2005distracted, walter2021cognitive}. Under such heightened load not only do our attentional resources suffer, but it also affects where and how we direct our visual attention \cite{lavie2005distracted}. As \changed{our focus on a task intensifies}, `tunnel vision' may occur, impeding our ability to perceive changes in peripheral vision — resembling \textit{peering through a tunnel}. This aligns with the `general interference' model, proposing that in this heightened state of focus, the boundary between attentive and inattentive zones becomes unclear \cite{young2012cognitive, ringer2016impairing}. In the case of user interfaces, this is especially relevant when multi-tasking, as attentional tunneling can reduce sensitivity to visual changes \cite{lee2007visual} and cause us to overlook important information. \changed{Additionally, prior work} studied the effect of cognitive load on visual attention in AR and VR settings \cite{rantanen1999effect,saint2020red,baumeister2017cognitive,lee2023visual, wickens2009attentional}. \changed{However,} its exploration in the context of the effectiveness of visual cues, in particular while looking at webpages, and its impact on saliency prediction is limited. In this study, we, for the first time, explore these phenomena together to determine how they impact the effectiveness of visual cues and together modulate the viewing behaviour.
\subsection{Visual Saliency Prediction: Datasets and Models}
Visual saliency prediction helps forecast image regions likely to attract human attention, treating it as a binary segmentation task by assigning a saliency score to each pixel based on its attention-grabbing potential \cite{liu2010learning}. Initially, bottom-up saliency models relying on low-level image features were used, but with improving deep learning approaches, data-driven solutions \cite{wang2018deep, leiva2020understanding} have come to the forefront. Meaningful saliency prediction, thus, relies on expansive eye-tracking datasets capturing human gaze patterns and computational models capable of learning from this data.

While extensive research has explored salience in natural scenes \cite{jiang2015salicon,judd2009learning}, gaze patterns in UI contexts, marked by a top-left fixation bias and a focus on text over images, present unique challenges  \cite{leiva2020understanding,jiang2023ueyes}. Although existing datasets, FiWI \cite{shen2014webpage} and UEyes \cite{jiang2023ueyes} \changed{helped offer the above insights into user gaze behaviour on an array of UIs, more comprehensive research on it remains sparse especially none that consider} the impact of cognitive load. Even state-of-the-art saliency models based on VGG-16 ~\cite{simonyan2014very,kruthiventi2017deepfix} and ResNet~\cite{he2016deep,liu2018deep} trained only on datasets with static images of natural scenes, fall short when handling dynamic UI interactions; and so do video-based saliency models \cite{min2019tased, chang2021temporal}, \changed{with  complex architectures designed for long videos with high frame rates, which appear excessive} for predicting dynamic but subtle changes in UIs (e.g., notifications appearing). 
%, prompting the consideration of a simpler model in our study. 
This gap in existing models because of the absence of any data on how dynamic changes influence bottom-up attention and the effect of cognitive load emphasizes the need for the use of simpler models in studying attention dynamics in dynamic UIs. This paper, accompanied by our new dataset, \dataset{}, which provides eye-tracking data from users looking at dynamic highlights on UIs, under the demands of varying cognitive load, lays the foundation for advanced computational models to predict user attention in real-world UI interactions.

\section{User Study}
%TODO add an intro to the study, referring to the research questions 
\changed{The goal of our empirical study is to understand how users attention is influenced by dynamically appearing highlights and task-induced cognitive loads.} Thus, our study manipulates two variables: the highlighting (`Highlight') of UI elements in the stimuli presented to participants, and the cognitive load (`Cognitive Load') of the secondary task which participants performed while visually inspecting the UIs.

% to ADD FINAL Values
\subsection{Participants}\label{sec:partcipants}
A pilot study was conducted with $5$ volunteers to refine our experiment procedure, interface and materials. For the main study, we recruited $30$ participants through \changed{word-of-mouth referrals and university ads} but had to exclude data from $3$ due to poor data quality. Our final participant group included $27$ individuals ($M_{Age} = 25.4 \pm 4.55$; \textit{Range:} $20-37$), consisting of $12$ female, $14$ male, and $1$ non-binary participants representing $13$ different nationalities. None of the participants, including the $17$ with visual aids, had a history of colourblindness or epilepsy. Before obtaining their written consent, participants were provided with information about the study's objectives, their rights, and any potential risks. The study took $45$ minutes on average, and participants were paid $10$ EUR. 
% 10 €.
% do you need to mention (±5) minutes?
% The study was approved by the ethical review board of the authors' university. 

\subsection{Stimulus Design}\label{sec:materials}
\begin{figure}[t]
  \centering
\includegraphics[width=1\textwidth]{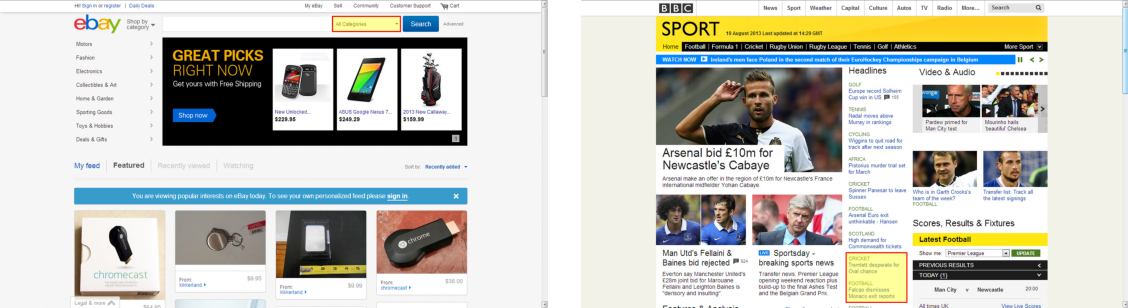}
  \caption{We controlled the size and location of highlighted areas to be equally distributed across images. Left: An example of a small ($11 cm^2$) highlight in location \textit{Q2} on a low clutter ($FC = 5.05$) UI. Right: A big $(33.3 cm^2$) highlight in \textit{Q4} on a complex UI ($FC = 7.1$). See \autoref{fig:probability highlight seen} and Supplementary (Supp.) Material for more details.}
  \label{fig:HT_Stimulus_Types}
\end{figure}

We selected webpages as a suitable proxy for user interfaces (UIs) to maintain research consistency, given their shared characteristics with most other UIs.\footnote{Webpages exhibit similar characteristics as most UIs -- complex, diverse, with numerous attention-grabbing elements. This choice helps maintain consistency with respect to visual angles etc. The terms -- interfaces, UIs, and webpages may be used interchangeably throughout this paper.} 
We chose two publicly available eye-tracking datasets, FiWI~\cite{shen2014webpage} and the more recent UEyes \cite{jiang2023ueyes} from which we selected 150 images, excluding overly simple designs (based on the Feature Congestion FC) Score \cite{rosenholtz2005feature}, as provided by \citet{oulasvirta2018aalto}), designs with a dark background, explicit content, and non-English texts. 
All images were resized to 2560 $\times$ 1440 pixels and padded with white borders to maintain the aspect ratio. 
On each webpage image, we selected a non-salient area to highlight, based on the fixation map provided by the respective dataset, i.e. a UI element that was never or seldom looked at by participants. Areas were highlighted by drawing rectangular boxes with a yellow background and a red border, as shown in \autoref{fig:HT_Stimulus_Types}. 
% We controlled the size and location of highlighted areas to be equally distributed across images, as described in the Supplementary Material. 
% (more details in \ref{Supplementary Material})

Based on these 150 images and selected areas, we generated three stimulus sets using different highlighting techniques: 
(1) \textsf{Absent}: 150 images of webpages as in the original datasets, (2) \textsf{Static}: the same images as in \textsf{Absent} but one area is permanently highlighted, (3)
    \textsf{Dynamic}: an animated \textit{gif} where an image from \textsf{Absent} was shown for three seconds after which the same image from \textsf{Static} was shown for two seconds, giving the impression of a dynamically appearing highlight on the webpage.

% to ADD FIGURE - FC SCORE, SIZE AND LOCATION

\subsection{Study and Task Design}\label{sec:task design}
Participants' primary task was to look at 150 images ``naturally as [they] would while surfing the web''. Stimuli were presented randomly over six blocks of 25 images each. Each image was displayed for $5s$, followed by a $2s$ blank screen interval. The study followed a within-subject design with two independent variables with three levels each:
    \textbf{Highlighting Technique (HT)} with \textsf{Absent}, \textsf{Static}, \textsf{Dynamic} and 
    \textbf{Cognitive Load (CL)} with \textsf{Absent}, \textsf{Low}, and \textsf{High}.
\textbf{HT} corresponds to the respective set of stimuli, as described above. Stimuli were randomly displayed so that each block contained stimuli from all Highlighting techniques.
\textbf{CL} denotes the secondary task participants were instructed to perform. Either it was absent, or, following prior work \cite{duchowski_index_pupillary, lindlbauer_context_aware}, participants were asked to count out loud from 0 onward in steps of 2, or backwards in steps of 7 starting at 800, \changed{inducing low and high cognitive load compared to \textsf{Absent} condition as confirmed by a post-study questionnaire and a statistical analysis of the pupil dilation (see Supplementary material for details)}.
At the beginning of each block, the instruction for counting was presented on the screen as well as verbally by the experimenter. Cognitive Load was balanced across participants and its order randomized across blocks.   
A Latin square design was used to create balanced lists of images so that each participant viewed each of the 150 unique webpages once and experienced all nine conditions. Accordingly, the design ensured that all webpages were looked at by at least three participants under all Highlighting Techniques \textbf{HT} and Cognitive Load \textbf{CL} conditions.

\subsection{Procedure and Apparatus}\label{sec:procedure}

After familiarizing themselves with the interface through two blocks  of practice trials (viewing stimulus images, not part of the main task, under \textsf{High} and \textsf{Low} CL), participants underwent a 9-point calibration on the Tobii Pro ETM software \cite{TobiiProLab}, immediately followed by a 9-point calibration verification on our custom interface.\footnote{\label{techspecs}Built using HTML, JavaScript (JS), and Flask v2.3.2; Python v3.8, tobii-research 1.10.1} After successful calibration with an average offset $<0.5^{\circ}$ ($\sim30\, \text{px}$), they proceeded to the main tasks, as described above, and finished with a post-study questionnaire (see suppl. material). Participants were comfortably seated in a stationary chair in a dimly-lit room, approximately 65-75 cm away from a 25.5-inch monitor with a resolution of 2560 $\times$ 1440 pixels. Raw eye gaze was captured with Tobii Pro Fusion (250 Hz), attached at the base of the monitor, and saved in CSV format as provided by Tobii SDK. After each block, participants took a short break during which the two-fold calibration was repeated.

% \textsuperscript{\ref{techspecs}}.
% The entire experiment was divided into six blocks, each containing 25 stimulus images. Each image was displayed for a duration of 5 seconds, followed by a 2-second blank screen interval [cite]. Before starting each block, participants were assigned a CL-inducing secondary task, which they had to perform throughout the entire block. Both the order of the secondary tasks and the order of the stimulus images were randomized to mitigate potential order effects. After each block, participants took a short break, and the two-fold calibration was repeated before starting the new block. % before starting the new block.

% \textit{REMINDER: Authors are required to include a privacy and ethics statement related to the study conducted in the \textbf{methodology} section}

\subsection{Data Processing}\label{sec:data processing}
For each timestamp, we save gaze points as $(x, y)$ in screen coordinate space. For data processing, we followed recommendations suggested by \citet{feit17}. Single outliers are identified by comparing each gaze point with its preceding and succeeding points. If the deviation from these nearby points exceeds $1^{\circ}$ in x- or $1.2^{\circ}$ in y- direction, the gaze point is marked as noise and corrected by replacing it with the median of its neighbours. Inter-sample velocity is then calculated and utilized to identify potential saccades, allowing us to apply a Weighted On-Off filter with Gaussian kernel smoothing \cite{spakov_comparison} to the raw gaze data. This helps stabilize high-frequency data without introducing any saccade-delays or latencies. The smoothed gaze data exhibits a higher level of accuracy and precision \footnote{Average accuracy $= 18.26 px (0.3^{\circ})$ and average precision $= 8.44 px (0.15^{\circ})$} compared to raw gaze data and is used to extract fixations. We implemented an algorithm for fixation detection, as proposed by Kumar et al. \cite{kumar2008improving, kumar2007guide} using Python. In this way we could retain 99.3\% of the gaze data, excluding trials with data loss $>25\%$ and gaze time on screen (GTS) $<80\%$ (\citet{hvelplund2014eye} define \tiny
\textbf{GTS} = $\left(\frac{\text{total fixation duration}}{\text{total task time}}\right) \times 100$ \normalsize) as per-stimulus data quality metrics. %\footnote{Note: Excluding the gaze data of the 3 participants that we dropped.} 
The processed fixation data was then used to generate fixation maps by smoothing them with a 2D Gaussian filter. We extract \emph{salient regions} that were looked at most, by  segmenting these heatmaps and extracting information about the salient regions' location, dimensions, and intensity.\footnote{Using OpenCV's v4.7 connectedComponentsWithStats}

% TODO: COMBINE ALL FOOTNOTES regarding technical package specs INTO 1.
% \textbf{Raw} Data: Avg Accuracy $= 21.26 px (0.38^{\circ})$ \& Avg Precision $= 13.46 px (~0.24^{\circ})$

% The processed fixation data was then used to generate heatmaps (smoothed with a 2D Gaussian filter), aggregated across all three CL conditions for each Highlight type -- resulting in a total of 450 Ground Truth (GT) heatmaps, with 150 heatmaps per Highlight type
% % \footnote{ We can achieve further granularity, by only averaging over per Highlight per CL. This was not employed for our analysis but was employed during saliency modelling} 
% , each representing the overall fixation data from nine participants. To identify regions that were salient for most users, we  segment the GT maps and extract information about the regions' location, dimensions, and intensity \footnote{Using OpenCV's connectedComponentsWithStats v4.7)}.
\section{Findings}

%TODO: i NEED to completely update change and rewrite 4.1  with updated modle selection and contrast coding%
% ALSO, I will change the colour of LOW CL from pink to a shade of yellow -- will do this at the end.

\subsection{Data Analysis}\label{sec:dataanalysis}

We use generalized linear mixed-effects models (GLMM) to analyze the eye-tracking data from our complex experiment design, because it allows us to account for random effects due to individual variability, webpage designs, or learning effects, and is particularly suited for non-normal data and our repeated measures design (\cite{catrysse2018learning, silva2022using}).
We include a brief description of the eye-tracking measures (response variables) in the GLMMs for each part, including their observed distribution (i.e. corresponding distribution family). We assume a log link between predictors and response variables.

For Section \ref{sec:ImageView4.2} both the HT and CL variables were encoded with the three levels using Helmert coding. For HT, the first comparison is that between \textsf{Absent} highlighting and the mean of \textsf{Static} and \textsf{Dynamic} highlighting. The second comparison is between \textsf{Static} and \textsf{Dynamic} highlighting. For CL, the model first compares no load to the mean of \textsf{Low} and \textsf{High} load. Its second comparison is between \textsf{Low} and \textsf{High} load. Aside from the main effects of HT and CL, we also included their interaction. When we noticed significant interaction effects from the GLMM analysis, \changed{Bonferroni-corrected post-hoc pairwise comparisons were performed}. For Sections \ref{sec:noticeabilityofAOIS} and \ref{sec:HRvsSRAnalysis} we encoded HT with 2 levels \textsf{Static} vs. \textsf{Dynamic} using sum coding; allowing us to report the direct comparison between the means of these two levels. In addition to these fixed effects, each regression model included random intercepts by participants, images, and block index, with random slopes for CL and HT by participant. We conducted our analyses in R \cite{r_manual} with \texttt{lme4} \cite{bates_lme4}, \changed{which yields \textit{t values} for the Gamma distribution and \textit{z values} for other (Poisson or Binomial) distributions}.

\subsection{Viewing Behaviour of Webpages}\label{sec:ImageView4.2}

% \begin{figure}
%     \centering
%     \vspace{-\baselineskip}
%     \begin{minipage}[b]{0.3\textwidth}
%         \centering
%         \includegraphics[width=\textwidth]{figures/plot1.png}
%         % \caption{TotalNumFixations}
%     \end{minipage}
%     \hfill
%     \begin{minipage}[b]{0.3\textwidth}
%         \centering
%         \includegraphics[width=\textwidth]{figures/MFD.png}
%         % \caption{MeanFixationDuration}
%     \end{minipage}
%     \hfill
%     \begin{minipage}[b]{0.3\textwidth}
%         \centering
%         \includegraphics[width=\textwidth]{figures/plot3.png}
%         % \caption{Num_SalientRegions}
%     \vspace{-\baselineskip}
%     \end{minipage}
%     \caption{Differences in viewing behaviour across the analyzed conditions. Left: The total number of fixations on the images. Middle: The average duration of fixations on the images in $ms$. Right: The number of salient regions as given by participants visual attention. The white dots and numbers represent the Mean, the black bars the median.}
%     \label{fig:ViewingImage}
%     \vspace{-\baselineskip}
% \end{figure}

\begin{figure}[ht]
    \centering
    \begin{subfigure}[b]{0.28\textwidth}
        \centering
        \includegraphics[width=\textwidth]{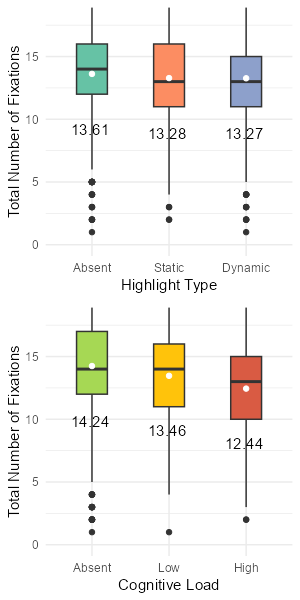}
        \caption{The total number of fixations on the images.}
    \end{subfigure}
    \hfill
    \begin{subfigure}[b]{0.28\textwidth}
        \centering
        \includegraphics[width=\textwidth]{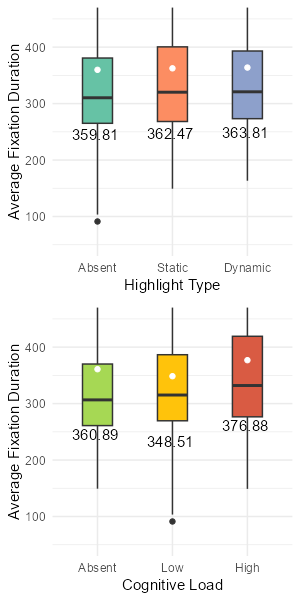}
        \caption{The average duration of fixations on the images in $ms$.}
    \end{subfigure}
    \hfill
    \begin{subfigure}[b]{0.28\textwidth}
        \centering
        \includegraphics[width=\textwidth]{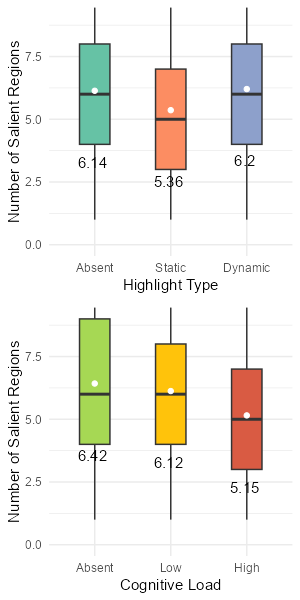}
        \caption{The number of salient regions as given by participants' visual attention.}
    \end{subfigure}
    \vspace{-\baselineskip}
    \caption{Differences in viewing behaviour across the analyzed conditions. White dots and numbers are the Mean, black bars the median.}
    \label{fig:ViewingImage}
    % \vspace{-\baselineskip}
\end{figure}
% \vspace{-\baselineskip}

We first wanted to understand the joint impact of HT and CL on the participant's viewing behaviour of the full webpage. We analyzed our data by computing the following metrics (name, description, distribution):
\vspace{0.2em}
\begin{table}[h]
\centering
\vspace{-\baselineskip}
\fontsize{8.5}{10}\selectfont
\begin{tabular}{lll}
\hline
\textbf{Number of Fixations} & Number of Fixations in an AOI or the overall image & Poisson\\ 
\textbf{Average Fixation Duration} & Average duration of Fixations in an AOI or the whole image, in $ms$ & Gamma \\ 
\textbf{Number of Salient Regions} & Overall number of regions of interest & Poisson \\
\hline
\end{tabular}
\vspace{-\baselineskip}
% \vspace{0.2em}
\label{table:OverallImageView}
\end{table} 

% \begin{table}[h]
% \centering
% \begin{tabular}{p{3.5cm}p{7cm}p{3cm}}
% \toprule
% \textbf{Category} & \textbf{Description} & \textbf{Distribution} \\
% \midrule
% Number of Fixations & Number of fixations in an AOI or the overall image & Poisson \\ 
% Average Fixation Duration & Average duration of fixations in an AOI or the whole image, in $ms$ & Gamma \\ 
% Number of Salient Regions & Overall number of regions of interest & Poisson \\
% \bottomrule
% \end{tabular}
% \end{table}

We found that participants made fewer but longer fixations in the presence of highlighting and with increasing cognitive load. 
\autoref{fig:ViewingImage} gives an overview of the results and specific numbers, which we discuss in the following. 

The total number of fixations across the images was largest in the case of \textsf{Absent} HT, as confirmed by the  GLMM analysis ($z = 3.732$, $p < 0.001$). However, the type of highlight (\textsf{Static} HT vs. \textsf{Dynamic} HT) did not show a significant effect ($z = 3.340$, $p = 
 \changed{0.79}$).
The number of fixations was also affected by CL with GLMM analysis revealing a highly significant \changed{decrease in fixations from} \textsf{Absent} CL to the presence of any ($z = 3.732$, $p < \changed{0.001}$). Fixations further decreased significantly from \textsf{Low} to \textsf{High} CL ($z = 3.577$, $p < \changed{0.001}$).
%We did not found a statistically significant interaction effect between HT and CL.  Post hoc pairwise comparisons revealed that differences in Highlight didn't significantly impact the pronounced effect of Cognitive Load, especially when comparing Absent - High load conditions ($p < 0.005$) and Low - High ($p < 0.05$) for all three levels of Highlight.

As the number of fixations reduced, the average duration of fixations across the images increased. Participants fixated longest under \textsf{High} CL compared to \textsf{Low} and \textsf{Absent} CL, where the median of fixation duration was $25 ms $ shorter in the latter case. The GLMM model shows a significant difference between \textsf{High} CL and \textsf{Low} CL ($t = -2.119$, $p = 0.034$). However, with the Helmert coding, no significant effect is found when comparing \textsf{Absent} CL vs. presence of CL (Low and High) ($t = -0.716$, $p = \changed{0.47}$).
\autoref{fig:ViewingImage} indicated an increase in the fixation duration over the image in the presence of a visual highlight of $10 ms$ compared to \textsf{Absent} HT. However, this difference was not significant ($t = -1.672$, $p = \changed{0.095}$) and no significant interaction effect was found between CL and HT. 

As a result of the observations above we see that participants explore fewer parts of the images when under cognitive load and when a particular region of the image was highlighted. In the \textsf{Absent} CL condition, the number of salient regions in an image (extracted from the corresponding heatmap as described in Section \ref{sec:data processing}) was the largest ($z = 5.422$, $p<0.0001$). Comparing \textsf{High} CL versus \textsf{Low} CL, participants looked at one region less, a statistically significant difference ($z = 5.390$, $p<0.0001$). \autoref{fig:ViewingImage}  also shows the effect of the presence of a highlight in the stimulus image which significantly ($z = 2.165$, $p = \changed{0.0303}$) reduced the number of salient regions from an image without any highlights (\textsf{Absent} HT), compared to images with \textsf{Static} HT and \textsf{Dynamic} HT. However, a significant \changed{increase in the number of salient regions is also observed} between \textsf{Static} HT and \textsf{Dynamic} HT ($z = -5.023$, $p<0.0001$).  

%% We might want to move this summary / interpretation to the discussion to not repeat too much there?
In summary, we observed that the highlighting technique (HT) affected how participants viewed the webpage. The presence of a highlight led to a more focused gaze behaviour, resulting in fewer and longer fixations, compared to the \textsf{Absent} HT condition. The type of highlight (\textsf{Static} or \textsf{Dynamic} HT) did not make a difference in the number or duration of fixations. Nevertheless, the number of explored regions was much larger in the \textsf{Dynamic} HT, than the \textsf{Static} HT, where participants broadly explored the webpage before the highlight appeared after three seconds. 
When under \textsf{High} CL, participants looked at fewer regions in the image, making fewer and longer fixations. 
In the case of \textsf{Low} CL, the effect was not as strong and participants were able to explore a similar number of different areas of the webpage. Interestingly, no interaction effect was found between HT and CL.

\subsection{Noticeability of Highlights}
\label{sec:noticeabilityofAOIS}
We next wanted to understand what factors influenced participants' gaze behaviour in detecting highlighted areas and how they affected their bottom-up attention, based on the following metrics (name, description, distribution):

\vspace{0.6em}
\begin{table}[h]
\centering
\vspace{-\baselineskip}
\fontsize{8}{9.5}\selectfont
\begin{tabular}{lll}
\hline
\textbf{AOI hits} \textit{\fontsize{6}{10}\selectfont(here, AOI = highlighted region)}& A binary value indicating whether any fixation fell within a spec. AOI & Binomial\\
\textbf{Time to First Fixation} & Time from highlight onset to initial fixation in highlighted region & Gamma \\ 
\textbf{Distance from Last Fixation} & Distance from last fixation to fixation in highlighted region & Gamma \\ \hline
\end{tabular}
\vspace{-\baselineskip}
\label{table:BUAMetrics}
\end{table}

% \begin{table}[h]
% \centering
% \fontsize{7.3}{9}\selectfont % Set custom font size for the entire table
% \begin{tabular}{lll}
% \hline
% \textbf{AOI hits} \textit{(here, AOI = highlighted region)} & A binary value indicating whether any fixation fell within a specified AOI & Binomial \\
% \textbf{Time to First Fixation} & Time from highlight onset to initial fixation in highlighted region & Gamma \\ 
% \textbf{Distance from Last Fixation} & Distance from last fixation to fixation in highlighted region & Gamma \\ 
% \hline
% \end{tabular}
% \label{table:BUAMetrics}
% \end{table} 

% To Change subsection heading %
\subsubsection{Highlight and Cognitive Load on AOI}
\label{sec:AOI_hit}
\begin{figure}
    \centering
    \begin{minipage}{0.35\textwidth}
        \centering
        \includegraphics[width=\textwidth]{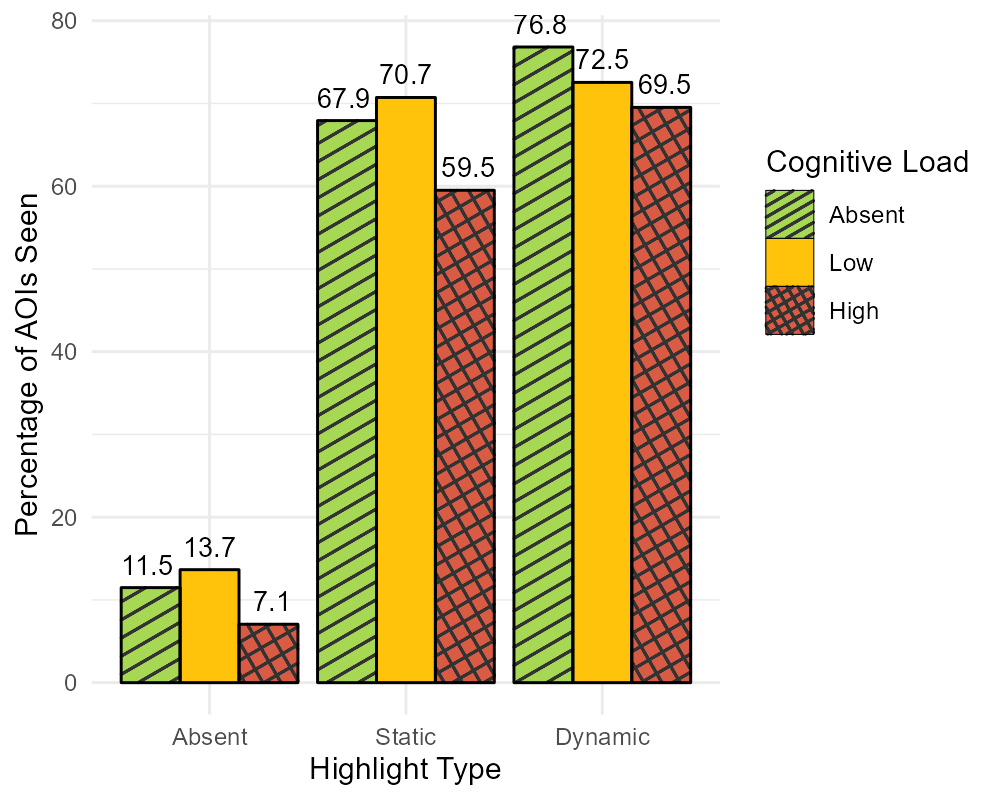}
        \caption{The percentage of fixated AOIs (highlighted areas) in each condition.}
        \label{fig:aoi}
    \end{minipage}
    \hfill
    \begin{minipage}{0.62\textwidth}
        \centering
        \includegraphics[width=\textwidth]{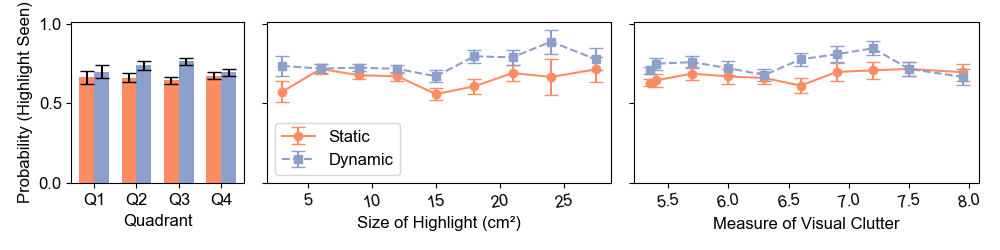}
        \caption{The probability that a highlighted area is fixated across different characteristics of the stimulus. We measure the visual clutter of the stimulus using the Feature Congestion metric, while size is operationalized as the area covered by the highlight in cm squared; and location is categorized by dividing the stimulus into four quadrants: Quadrant 1 (Q1) represents the top-left, Q2 the top-right, Q3 signifies the bottom-left, and Q4 the bottom-right.}
        \label{fig:probability highlight seen}
    \end{minipage}
    % \caption{AOIs Seen : TODO Describe and Change plot 1's height}
\end{figure}

\autoref{fig:aoi} summarizes the AOI hits across conditions. \textsf{Static} HT triggered a substantial $55.2\%$ rise in AOI hits, making a previously inconspicuous region noticed by participants. \textsf{Dynamic} HT was the most attention-grabbing with 62.2\% more AOI hits ($z = -4.031, p < 0.0001$).
Participants were hindered in their ability to notice an AOI when under \textsf{High} CL. There was a 13.1\% decrease in AOI hits compared to when under \textsf{Low} CL, a statistically significant difference ($z = 4.675, p < 0.0001$). However, the difference in AOI hits between \textsf{Low} CL and \textsf{Absent} CL was a minimal 0.4\%. While GLMM analysis confirmed a significant effect in the presence of any CL compared to \textsf{Absent} CL,  pairwise post-hoc tests could not find a significant difference between \textsf{Absent} HT and \textsf{Static} HT ($z = -0.93, p = 1.00$). Interestingly, when the stimulus image had either \textsf{Absent} HT or \textsf{Static} HT, participants attended to 3.4\% more AOIs while counting up in steps of 2 compared to freely looking at the webpage (see \autoref{fig:aoi}). 
Accordingly, in the \textsf{Static} HT condition, there is a significant difference between \textsf{Low} CL and \textsf{High} CL ($z = 3.464, p = 0.0016$). However, this difference is not observed in the \textsf{Dynamic} HT condition ($z = 0.987, p = \changed{0.97}$). 

To summarize, our findings indicate that \textsf{Dynamic} HT was most effective in attracting participants' attention to otherwise unnoticed areas. \textsf{High} CL markedly reduced the number of hit AOIs in the \textsf{Static} HT condition. However, \textsf{Dynamic} HT counteracted this effect and participants detected AOIs in both CL conditions equally well. 

% Interaction effects present: significant. BEHAVIOUR OF LOW CL --to Discuss; how to interpret %

\subsubsection{Stimulus Characteristics}

In Section \ref{sec:materials} we described our stimulus curation, balancing various characteristics: visual complexity of the image, highlighted region size, and location. We wanted to explore if these variables strongly influenced the probability of participants noticing a highlighted region. Fig. \ref{fig:probability highlight seen} suggests that none of these variables significantly influenced the highlight detection probability in our experimental setting. A GLMM model with these factors as the dependent variables and a crossed-random effects was fit. The results ($p > 0.1$ for each) affirm that none of them played a consistently significant role in the noticeability of a highlight, validating the reliability of our experimental design and emphasizing the primary role of controlled factors in explaining any variability.

% \begin{figure}
%     \centering
%     \includegraphics[scale=0.59]{figures/probability highlight seen.png}
%     \caption{We measure visual complexity of the stimulus using the Feature Congestion metric, while size is operationalized as the area covered by the highlight in cm squared; and location is categorized by dividing the stimulus into four quadrants: Quadrant 1 (Q1) represents the top-left, Q2 the top-right, Q3 signifies the bottom-left, and Q4 the bottom-right.}
%     \label{fig:probability highlight seen}
% \end{figure}

\subsubsection{Dynamic Highlight and Bottom-up Attention}\label{sec:Bottom-UpAttention}

% \begin{figure}
%     \centering
%     \includegraphics[width=0.8\linewidth]{figures/TtFF.png}
%     \includegraphics[width=0.8\linewidth]{figures/DfLF.png}

%     \caption{Comparing noticeablity of highlighted regions under experimental conditions and their fitted interaction plots. Row 1 compares Time to First Fixation in $ms$, Row 2 the Distance from the Last Fixation (previous focus of attention) in $px$}
%     \label{fig:BottomUpAttention}
% \end{figure}

\begin{figure}[ht]
    \centering
    \begin{subfigure}[b]{0.8\linewidth}
        \centering
        \includegraphics[width=\linewidth]{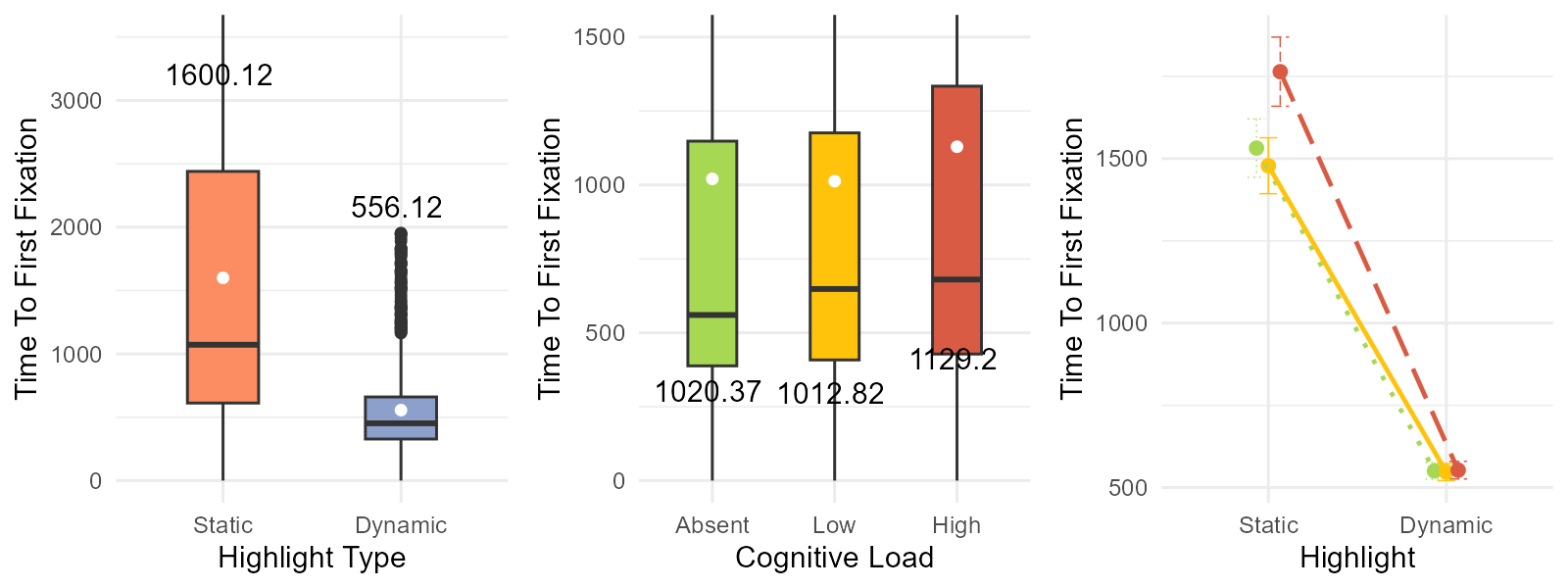}
        % \caption{Comparing Time to First Fixation in $ms$}
        \label{subfig:time_to_first_fixation}
    \end{subfigure}
    \vspace{-\baselineskip}
    \begin{subfigure}[b]{0.8\linewidth}
        \centering
        \includegraphics[width=\linewidth]{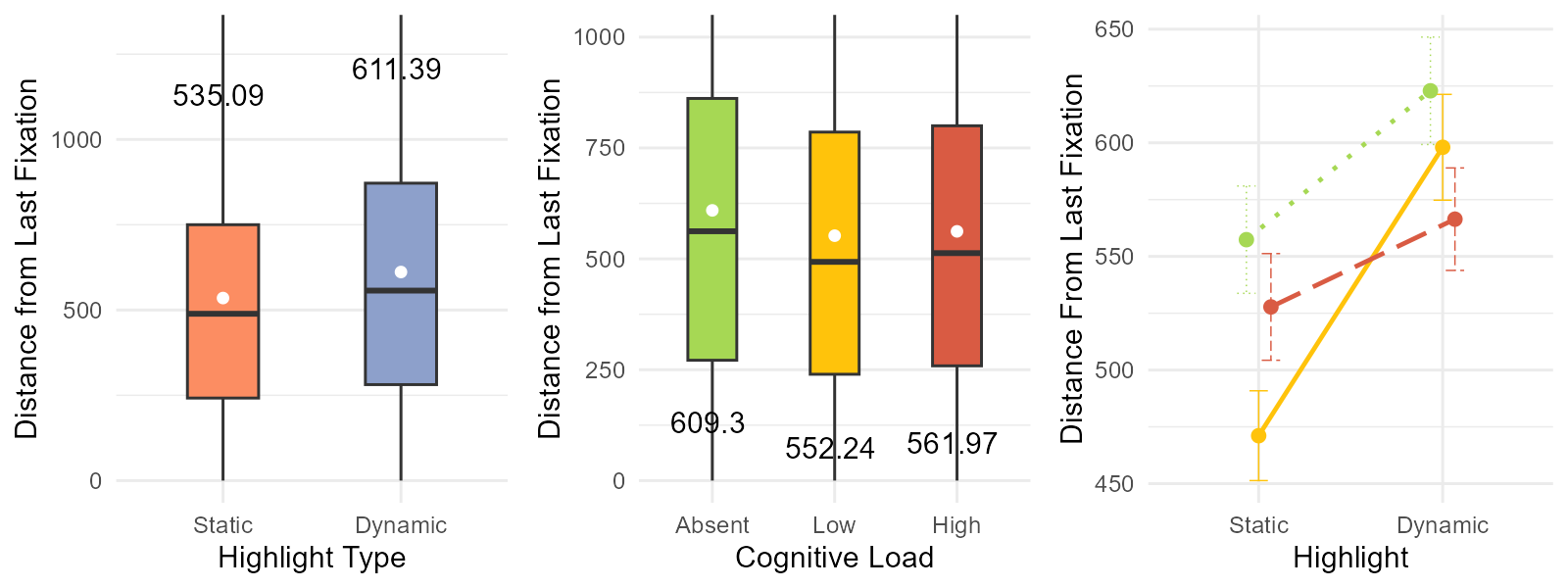}
        % \caption{Comparing Distance from the Last Fixation (previous focus of attention) in $px$}
        \label{subfig:distance_from_last_fixation}
    \end{subfigure}
    \caption{Comparing noticeablity of highlighted regions under experimental conditions and their fitted interaction plots. Row 1 compares Time to First Fixation in $ms$, Row 2 the Distance from the Last Fixation (previous focus of attention) in $px$.}
    \label{fig:BottomUpAttention}
\end{figure}

% \begin{figure}[b]
%     \centering
%     \vspace{-\baselineskip}
%     \begin{minipage}[b]{0.3\textwidth}
%         \centering
%         \includegraphics[width=\textwidth]{figures/plot4.png}
%         % \caption{TtFF}
%     \end{minipage}
%     \hfill
%     \begin{minipage}[b]{0.3\textwidth}
%         \centering
%         \includegraphics[width=\textwidth]{figures/plot5.png}
%         % \caption{DistanceFromLastFixation}
%     \end{minipage}
%     \hfill
%     \begin{minipage}[b]{0.3\textwidth}
%         \centering
%         \includegraphics[width=\textwidth]{figures/InteractionTtFFF.png}
%         % \caption{InteractionTtFFF}
%     \end{minipage}
%     \caption{Comparing noticeablity of highlighted regions under experimental conditions and their fitted interaction plots. Row 1 compares Time to First Fixation in $ms$, Row 2 the Distance from the Last Fixation (previous focus of attention) in $px$}
%     \vspace{-\baselineskip}
%     % TO REGENERATE PLOTS - ROW 1 TtFF & Row 2 DlFF
%     \label{fig:BottomUpAttention}
% \end{figure}

In Section \ref{sec:AOI_hit}, \textsf{Dynamic} HT proved to be more effective in attracting participants' attention towards a UI region than \textsf{Static} HT. To better understand this, we compared \textsf{Dynamic} vs \textsf{Static} HT on how quickly and from how far highlighted areas were discovered.

\autoref{fig:BottomUpAttention} gives an overview of our findings. Regions with \textsf{Dynamic} HT were not only discovered $2.9$ times quicker than those with \textsf{Static} HT, but \textsf{Dynamic} HT also attracted participants' gaze from a greater distance. We found that the Time to First Fixation \changed{(TtFF)} was significantly \changed{shorter} for \textsf{Dynamic} compared to \textsf{Static} HT ($t = -19.780$, $p < 0.0001$), and the Distance from the Last Fixation (DfLF) was significantly \changed{greater} for \textsf{Dynamic} HT ($t = 4.586$, $p < 0.0001$).
% We found a significant difference between \textsf{Static} and \textsf{Dynamic} HT for the Time to First Fixation (TtFF) ($t = -19.780$, $p < 0.0001$, \changed{shorter for \textsf{Dynamic}}) and the Distance from the Last Fixation (DfLF) ($t = 4.586$, $p < 0.0001$, \changed{longer for \textsf{Dynamic}}).

Participants' first fixation on the highlighted region occurred earliest under \textsf{Absent} CL, later under \textsf{Low} CL, and on average $120ms$ later under \textsf{High} CL \changed{compared to \textsf{Absent} CL}. While the TtFF in the presence of CL (both \textsf{Low} + \textsf{High}) did not differ significantly from \textsf{Absent} CL ($t = -0.777, p = 0.437$), the presence of \textsf{High} CL had a statistically significant ($t = 2.391, p = 0.0168$) effect on TfFF. We also observed significant interaction effects ($t = 2.142, p = 0.0322$), and post-hoc analysis revealed that despite the presence of \textsf{High} CL, there were no statistically significant differences in the TtFF when the region was dynamically highlighted; but when the region had \textsf{Static} HT there were significant differences in the TtFF between \textsf{Low} vs. \textsf{High} CL ($z = -3.128$, $p = 0.0053$) and \textsf{Absent} vs. \textsf{High} CL ($z = -2.469$, $p = 0.0407$).

In \autoref{fig:BottomUpAttention} we see that participants \changed{were} able to travel the longest distance from their last fixation to the highlighted region when not under any cognitive load. Surprisingly, DfLF under \textsf{Low} CL was a median of $20px$ lesser than that of \textsf{High} CL. GLMM analysis revealed that the observed difference in DfLF between when participants were not under any CL compared to when it was present (both \textsf{Low} + \textsf{High}) was statistically significant ($t = 2.950, p = \changed{0.003}$). A significant interaction effect ($t = 2.336, p= 0.0194$) can also be seen in \autoref{fig:BottomUpAttention}, and post-hoc analysis revealed significant differences between \textsf{Absent} vs. \textsf{Low} CL conditions in the presence of \textsf{Static} HT ($z = 3.342, p = 0.002$). As with TtFF, there were no significant differences across CL conditions in the presence of \textsf{Dynamic} HT.

Putting the above findings together we can explain the effectiveness of \textsf{Dynamic} HT is largely due to it triggering participants' bottom-up attention mechanisms by mimicking `movement'. It led to a quicker, more urgent detection, showcasing attention-grabbing abilities even outside participants' foveal vision. The absence of any significant differences in TfFF and DfLF even between \textsf{Absent} vs. \textsf{High} CL in regions with \textsf{Dynamic} HT cements its ability to remain noticeable, coping well even when participants are under CL. %The lower DfLF under \textsf{Low} CL might indicate a strategic scanning behaviour worthy of discussing in \ref{sec:Discussion}.

% TODO Change Titles
\begin{figure}[b]
    \centering
     \begin{minipage}[b]{0.38\linewidth}
        \centering
        \includegraphics[width=\linewidth]{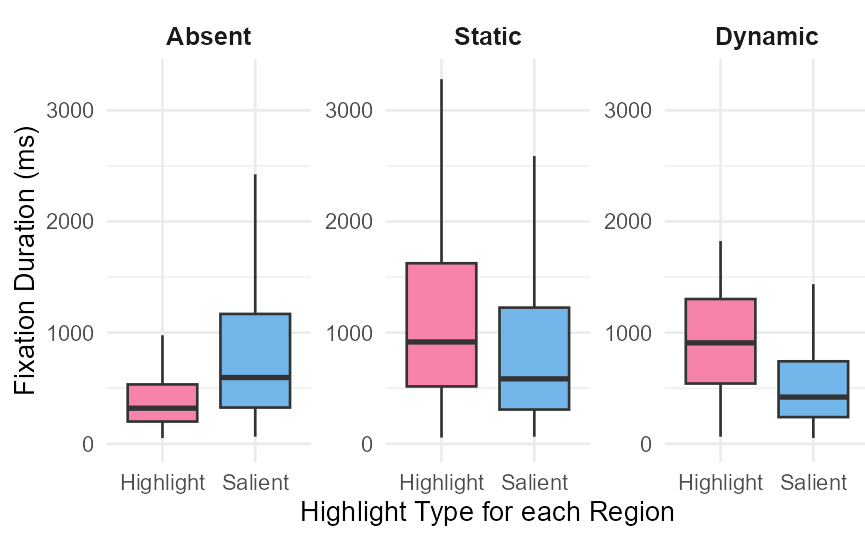}
        \caption{\changed{Fixation Duration on the Highlighted Region (pink) compared to the most ``naturally'' Salient Region (blue) across HT conditions.}}
        \label{subfig:fixation_duration}
    \end{minipage}
    \hspace{12pt}
    \begin{minipage}[b]{0.55\linewidth}
        \center
        \includegraphics[width=\linewidth]{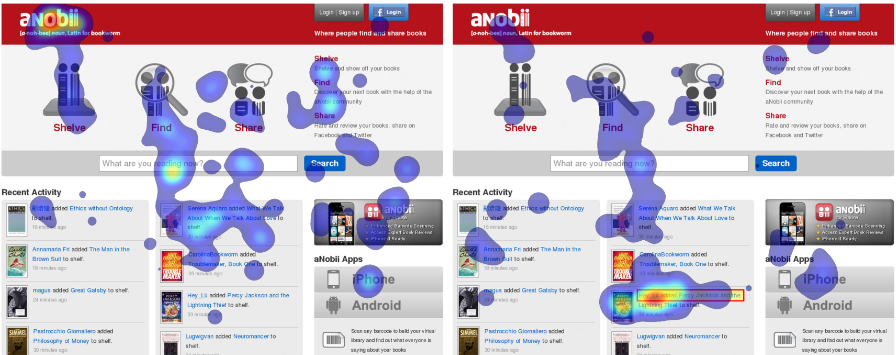}
        \caption{\changed{Example Fixation map of participants viewing a webpage in the \textsf{Absent} HT (left) compared to \textsf{Static} HT condition. The highlighted area attracts more attention than any other ``naturally'' salient region on the webpage.}}
        \label{subfig:heatmaps}
    \end{minipage}
    \label{fig:ShiftingAttention2}
\vspace{-\baselineskip}
\end{figure}

\subsection{Understanding Viewing Behaviour of Highlighted Regions}\label{sec:HRvsSRAnalysis}
% \section{Shift in Attention}
\changed{Finally, we wanted to explore whether a \emph{highlighted region} was looked at differently from other \emph{``naturally'' salient regions} of a webpage, such as titles, pictures with faces, buttons, etc. 
Therefore, we compared the Fixation Duration (see Section ~\ref{sec:ImageView4.2}) on a highlighted region to the duration on the otherwise most salient region of each webpage, across the three HT conditions. To determine the most salient region, we created fixation maps, by aggregating fixations from nine participants across all CL conditions and extracted salient regions, as described in Section \ref{sec:data processing}. Using OpenCV, we determined the most salient region as the one that was not the highlighted region and which had the highest intensity in the fixation map, indicating that it was fixated on the longest.}

%We ranked (see Supp. Materials for details) all the regions that were looked at by participants in the stimulus image based on the degree of engagement with it \footnote{Using OpenCV we could extract the intensity of the region of the fixation heatmaps. $\uparrow$ intensity $= \uparrow$ engagement = more and longer fixations}. The best-ranked element (second-best ranked when HR was the best) is what we term the SR. 

\autoref{subfig:fixation_duration} shows that, while ``naturally'' salient regions were looked at longer than regions with no highlight (Absent HT), in the presence of a visual highlight,  these highlighted regions (AOIs with Static and Dynamic HT) were fixated on longer than other salient regions, nearly doubling Fixations Duration. We fit GLMM models using the same method described in Section \ref{sec:dataanalysis} adding sum-coded \textsf{Region} (i.e., compared means of highlighted vs salient regions) as another fixed effect along with HT and CL, with random slopes and intercepts by participants, images, and block index. The difference in Fixation Duration was statistically significantly \changed{longer} ($t =7.533$, $p<0.001$) on the highlighted regions across different cognitive loads, compared to other salient regions. There were no significant interaction effects of HT or CL with \textsf{Region}.

\changed{In summary, our findings suggest that highlighting causes participants to shift their attention away from other salient regions of a webpage. As illustrated in \autoref{subfig:heatmaps}, attention on regions with visual highlights significantly increased to an extent that participants engaged longer with the highlighted area than any other area on the webpage. As a result, they explored fewer parts of the webpage, as we already observed in Section~\ref{sec:ImageView4.2}).
}

\section{Saliency Prediction}
Our empirical results showed that the presence of a highlight and its dynamics significantly altered the viewing behavior of webpages, not just for the highlighted region but also on the overall image. The cognitive load of participants further influenced their attention on the image. 
So far, deep-learning based saliency models, predicting the users fixations from pixel-level data, have not considered these modulating factors. Our goal in this part is to evaluate the performance of existing models and compare them to an implementation based on our \dataset{} dataset.

In selecting a model for predicting user saliency under varying experimental conditions, we prioritized the balance between computational efficiency and predictive accuracy. Given that \dataset{} is the first saliency dataset that explicitly considers HT and CL, we only had limited data for fine-tuning existing models. Thus, we chose SimpleNet \cite{reddytidying}, characterized by its architectural simplicity,  %, to mitigate the risk of overfitting that arises from our dataset's limited sample size. 
which previously demonstrated proficiency in identifying salient features within data-rich visualizations \cite{shin2022scanner}.
%, aligning closely with the nature of our UI stimuli, and thereby ensuring robust saliency predictions across different cognitive loads and visual highlight conditions.
The SimpleNet architecture incorporates a Progressive Neural Architecture Search (PNAS) as its encoder component, chosen for its high efficacy. The model's implementation was carried out using Python 3.9 and the PyTorch v1.10.2 framework. The training process was conducted on a workstation equipped with an Intel i7-13700KF CPU, 32GB of RAM, and an NVIDIA GTX 4080 Ti GPU. For fine-tuning the SimpleNet model as illustrated in the following Sections, the input images were resized to a resolution of $256\times256$ pixels. We began training with a learning rate of \(1 \times 10^{-4}\) and scaled it down by a factor of 0.1 every five epochs. Our loss function combines KL, CC, and NSS metrics, as used by prior work \cite{reddytidying,shin2022scanner}. Optimization was executed via the ADAM algorithm.

\begin{table}[b]
    % \centering
    \begin{minipage}{0.52\textwidth}
        % \centering
        \fontsize{7}{9}\selectfont
        \setlength{\tabcolsep}{3pt}
        \begin{tabular}{lp{0.6\linewidth}l}
            \hline
            \textbf{Metric} & \textbf{Description} & \textbf{Range} \\
            \hline
            AUC & Measures the model's effectiveness as a binary classifier for fixations & $[0, 1]$ \\
            NSS & Reflects the average normalized saliency at fixation points & $[-\infty, +\infty]$ \\
            SIM & Assesses the match between the predicted saliency map and the actual fixation distribution & $[0, 1]$ \\
            CC & The linear relationship between the predicted and actual fixation maps & $[-1, 1]$ \\
            KL & Quantifies the difference between the predicted saliency distribution and the ground truth & $[0, +\infty]$ \\
            \hline
        \end{tabular}
        % \vspace{-\baselineskip}
        \caption{Metrics for evaluating saliency models. See \cite{bylinskii2018different} for details.}
        \label{tab:saliency_metrics}
    \end{minipage}%
    \begin{minipage}{0.48\textwidth}
        % \centering
         \fontsize{7.5}{9}\selectfont
        \setlength{\tabcolsep}{3pt}
        \changed{\begin{tabular}{lcccccc}
            \hline
            \textbf{HT} & \textbf{Model} & \textbf{AUC} & \textbf{CC} & \textbf{NSS} & \textbf{SIM} & \textbf{KL} \\
            \hline
            Static & SALICON & 0.819 & 0.466 & 1.094 & 0.468 & 1.126 \\
            & WEB & 0.835 & 0.532 & 1.269 & 0.505 & 0.889 \\
            & Fine-tuned & \textbf{0.859} & \textbf{0.601} & \textbf{1.440} & \textbf{0.524} & \textbf{0.789} \\
            \hline
            Dynamic & SALICON & 0.820 & 0.454 & 1.135 & 0.460 & 1.093 \\
            & WEB & 0.863 & 0.512 & 1.468 & 0.491 & 0.910 \\
            & Fine-tuned & \textbf{0.900} &\textbf{0.630} & \textbf{2.057} & \textbf{0.540} & \textbf{0.749} \\
            \hline
        \end{tabular}}
        \caption{Model Performance under different highlights.}
        \label{tab:highlight_conditions_perform}
    \end{minipage}
    \vspace{-\baselineskip}
% \vspace{0.2em}
\end{table}

In the following, we explain the implementation of our saliency models which are \changed{based on the SimpleNet architecture pre-trained on the SALICON dataset \cite{jiang2015salicon, shin2022scanner} and fine-tuned on subsets of the \dataset{} dataset to account for the differences in gaze behavior due to highlighting or cognitive load. We compare the performance of these \emph{Fine-tuned} models against two baselines: (1) \emph{SALICON} denotes the SimpleNet model pre-trained on the SALICON dataset \cite{jiang2015salicon, shin2022scanner} and  (2) \emph{WEB} denotes that the model is further fine-tuned on a subset of our \dataset{} dataset where CL and HT were both absent, to adapt the model to the domain (webpages). The two baselines allow to disentangle the added value through fine-tuning for the specific domain and for highlighting and cognitive load. We compare the models' performance on a set of well-established metrics \cite{bylinskii2018different} listed in \autoref{tab:saliency_metrics}.}  A higher score is better in all cases except the KL metric.

\subsection{Saliency Prediction under Different Highlighting Techniques}
%Our empirical findings highlight the differences in gaze behavior as a result of \textsf{Static} and \textsf{Dynamic} highlighting, compared to \textsf{Absent} highlighting. Therefore, 
\changed{We first explored how dynamic and static highlighting affects the predictive performance of models trained on gaze data without any highlighting (\emph{SALICON} and \emph{WEB}). We developed two models by fine-tuning the \emph{SALICON} model with observations from the \textsf{Dynamic} and \textsf{Static HT}. Therefore, we created fixation maps for each \textsf{HT} condition by aggregating the fixation points from all users across all \textsf{CL} conditions and smoothing them with a 2D Gaussian filter ($SD=35px$). The resulting dataset, \dataset\textunderscore{HT}, comprises 150 stimuli for each \textsf{HT} condition. For each model, we randomly split the corresponding data into a training set (80\%, 120 images) and a testing set (20\%, 30 images).
%% I don't think that explanation is needed since the text above explicitly states that we devlop two models here.
%Given Baseline2's prior fine-tuning on the no HT condition, our analysis specifically targets the static and dynamic HT conditions. This focus helps to identify the specific influences of the visual cues to saliency prediction.
}
% We maintained an 80\%/20\% training and testing split for each subdataset and test the pre-trained SimpleNet in the same part of the data.%, enabling the models to specialize in the unique saliency characteristics of each condition while avoiding overfitting.

A key challenge in training our own model was to capture the temporal dynamics of the highlight in the \textsf{Dynamic} HT condition. Our solution involved extracting a pair of images from each dynamic stimulus (with and without the highlight) and feeding them as stacked inputs into the network. We realized this through a modified input layer with a specialized channel reduction convolution layer. This enables the network to discern salient feature differences between the image pairs and thus process the temporal dynamics. More details are provided in the Supplementary material.

\changed{Table \ref{tab:highlight_conditions_perform} compares the performance of the two baseline models trained on data without any highlighting to the fine-tuned models in the \textsf{Static} and \textsf{Dynamic} conditions. For the Static HT condition, the fine-tuned model achieves a notable improvement in the Correlation Coefficient (CC) by approximately 28.97\%, and in the Normalized Scanpath Saliency (NSS) by about 31.71\% over the SALICON baseline. In the Dynamic HT condition, enhancements are particularly significant in the CC, with an increase of 38.77\%, and in the NSS metric, where the fine-tuned model shows an impressive 81.23\% improvement. 
We also evaluated our architecture for handling dynamic highlights. \autoref{table:saliency_dynamic} shows the superiority of inputting paired images compared to training only on the image with the highlight (highlighted input) or randomly choosing the image with or without the highlight during the training process (random input). }

\changed{Overall, our findings indicate, that a generic saliency model (\emph{SALICON}) cannot fully capture the impact of a (dynamically appearing) highlight on a user's attention, even when fine-tuned on the image domain (\emph{WEB}).} Qualitative examples are given in the supplementary material.

\begin{table}[t]
% \centering
\begin{minipage}[t]{0.48\linewidth}
    % \centering
    \setlength{\tabcolsep}{3pt} 
    \fontsize{7.7}{9}\selectfont
    \changed{\begin{tabular}{lcccccc}
        \hline
        \textbf{Highlight} & \textbf{Model Input} & \textbf{AUC} & \textbf{CC} & \textbf{NSS} & \textbf{SIM} &\textbf{KL}\\ 
        \hline
        & Highlighted & 0.879 & \textbf{0.633} & 1.811  & 0.536 & 0.761\\
        Dynamic & Random & 0.868 & 0.552 & 1.571 & 0.512 & 0.873\\
        & Pair & \textbf{0.900} & 0.630 & \textbf{2.057} & \textbf{0.540} & \textbf{0.749}\\ 
        \hline
    \end{tabular}}
    \caption{Comparison of different approaches for inputting the dynamic stimulus during fine-tuning of SimpleNet. Our pairing approach performs best.}
    \label{table:saliency_dynamic}
\end{minipage}
\hfill
\begin{minipage}[t]{0.45\linewidth}
    \centering
    \fontsize{7.8}{9}\selectfont
    \setlength{\tabcolsep}{3pt}
    \changed{\begin{tabular}{lcccccc}
        \hline
        \textbf{CL} & \textbf{Model} & \textbf{AUC} & \textbf{CC} & \textbf{NSS} & \textbf{SIM} & \textbf{KL} \\
        \hline
        Low & SALICON & \textbf{0.875} & 0.356 & 1.229 & 0.325 & 1.689 \\
        & WEB & 0.864 & 0.391 & 1.223 & 0.343 & 1.496 \\
        & Fine-tuned & 0.870 & \textbf{0.406} & \textbf{1.490} & \textbf{0.349} & \textbf{1.476} \\
        \hline
        High & SALICON & 0.879 & 0.349 & 1.301 & 0.326 & 1.603 \\
        & WEB & 0.880 & 0.417 & \textbf{1.546} & 0.353 & 1.432 \\
        & Fine-tuned & \textbf{0.887} & \textbf{0.440} & 1.411 & \textbf{0.367} & \textbf{1.411} \\
        \hline
    \end{tabular}}
    \caption{Model Performance under different cognitive loads.}
    \label{tab:cognitive_load_perform}
\end{minipage}
\vspace{-\baselineskip}
% \vspace{0.2em}
\end{table}

% \begin{table}[H]
% \centering
% \caption{Performance of SimpleNet under different highlight conditions}
% \label{tab:highlight_conditions}
% \begin{tabular}{lrrrrrrrrrrrrrrr}
% \hline
% \multicolumn{1}{l}{\textbf{Highlight}} & \multicolumn{1}{l}{\textbf{Model}} & \multicolumn{1}{l}{\textbf{KL}} & \multicolumn{1}{l}{\textbf{AUC}} & \multicolumn{1}{l}{\textbf{CC}} & \multicolumn{1}{l}{\textbf{NSS}} & \multicolumn{1}{l}{\textbf{SIM}} \\
% \hline
% \multirow{2}{*}{Absent} & P & 0.9668 & 0.8035 & 0.5323 & 1.3259 & 0.4913 & \\
% & F & 0.6288 & 0.8531 & 0.6941 & 1.7897 & 0.6215 & \\
% \hline
% \multirow{2}{*}{Static}  & P & 1.1863 & 0.7779 & 0.4306 & 1.4659 & 0.4488 & \\
% & F & 0.7428 & 0.9345 & 0.6854 & 1.8908 & 0.6325 & \\
% \hline
% \multirow{2}{*}{Dynamic} & P &1.2744& 0.7836 & 0.3824 &  0.9803 & 0.4303 & \\
% & F & 0.8212 & 0.8546 & 0.5723 &  1.4230 & 0.5125 & \\
% \bottomrule
% \end{tabular}
% \end{table}

\subsection{Saliency Prediction under Cognitive Load %Condition Saliency Prediction
}

%In our empirical findings, we observed fewer but longer fixations under cognitive loads, suggesting a need for saliency models to account for cognitive load. To investigate this, 
\changed{We then explored the importance of accounting for the presence of cognitive load by implementing individual models for the \textsf{High} and \textsf{Low} CL conditions. Paralleling our approach in the previous Section, we created fixations maps for each CL condition by aggregating fixation points across users in all HT conditions. The resulting dataset, \dataset\textunderscore{CT}, comprises 150 fixation maps for each CL condition. We fine-tune SimpleNet models on the 150 heatmaps of each condition, as before, specializing in predicting saliency as per the cognitive state of the user. }
%As above, we split the data with 80\% for training and 20\% for testing.

\changed{Table \ref{tab:cognitive_load_perform} compares the performance of the two baseline models trained on data without any secondary task (i.e. \textsf{Absent\ CL}) to the models fine-tuned on data from the \textsf{Low} and \textsf{High} CL conditions. The fine-tuned models outperform the baselines in most metrics. In the case of \textsf{Low} CL the NSS score stands out with the fine-tuned model showing a 21\% higher performance compared to both baselines. Interestingly, in other metrics the difference is much smaller and also the two baselines perform quite similarly. The \textsf{High} CL condition reveals a similar trend, where the fine-tuned model outperforms both baselines on most of the metrics, but only by a small margin. In this case, the \emph{WEB} model shows a clearer performance improvement compared to the \emph{SALICON} model. }

\changed{Compared to the HT condition, differences in predictive performance are not as large when accounting for \textsf{Low} or \textsf{High} CL induced by secondary tasks. Interestingly, both, the \emph{SALICON} model trained on natural images and the \emph{WEB} model fine-tuned on webpages perform similarly well in predicting saliency in this condition. This variation shows the complexities of saliency prediction across different scenarios: the presence of explicit visual variation such as visual highlighting of inputs significantly influences model effectiveness, illustrating the challenges of accurately capturing user attention shifts driven solely by cognitive load differences.}

\section{Discussion and Conclusion}\label{sec:Discussion}
To the best of our knowledge, this is the first study to explore the joint impact of both dynamic and static visual highlighting, and cognitive load on users' attention when viewing complex interfaces, such as webpages. In times where technology constantly competes for our attention, it is important to understand which mechanisms affect the distribution of this limited resource. 

We carefully designed a free-viewing task that presented participants with overall 150 webpage images in different conditions: in their original design, with a permanently highlighted area or a dynamically appearing highlight. In 4 of 6 blocks, participants performed a secondary task of counting out loud which induced either high or low cognitive load, as confirmed by participants' subjective feedback. We carefully designed the stimuli and performed a thorough statistical analysis using GLMMs to account for random effects, e.g. due to individual differences, stimulus designs, or learning. This enabled us to attribute the changes we observed in visual attention to the independent variables in our experiment: Highlighting technique and cognitive load. In particular, we made the following observations: 

\emph{Webpages are explored less, when specific content is highlighted}, as indicated by fewer and longer fixations over the image. However, the \textsf{Dynamic} highlight, appearing only after three seconds, allowed participants to attend a similar number of image regions as for the \textsf{Absent} condition.

\emph{Webpages are explored less, when users experience high cognitive load.} Confirming prior work~\cite{walter2022low}, participants made fewer and longer fixations under \textsf{High} CL, indicating increased processing time. This was not the case under \textsf{Low} CL where users explored a similar number of regions compared to the \textsf{Absent} condition, despite multitasking. 

\emph{Dynamic highlighting attracts attention efficiently even under high cognitive load}. Highlighted areas were attended faster and from a larger distance from the periphery, in the \textsf{Dynamic} HT condition. In particular, this was also the case under \textsf{high} cognitive load, which in the \textsf{Static} case led to a slower shift of attention.  

\emph{Highlighted regions are engaged more with than any other salient region}, as indicated by significantly more and longer fixations on the highlighted region, compared to the most looked at region of the original webpage. This indicates, that participants engaged with highlighted information more thoroughly.

\subsection{Shifting the Focus of Saliency Prediction to Consider Highlighting and Cognitive Load}
Our empirical findings motivated us to explore whether state-of-the-art saliency models can account for the impact of highlighting and cognitive load to users attention. 
We implemented condition-specific saliency models based on a  state-of-the-art architecture, SimpleNet, fine-tuned on our dataset and achieved superior performance across all experimental conditions, compared to a pre-trained model. 
In particular, we presented the first image-based saliency model that can deal with temporal changes in an interface due to dynamic highlighting. We achieved this by feeding paired images (with and without highlighting) as stacked inputs into the network during  training. This led to a significant improvement in predictive performance. 

Importantly, our results highlight the need for saliency models to
explicitly incorporate (temporal) highlight information, and \changed{potentially} tailor them to the specific cognitive state of the viewer, to achieve more accurate and robust saliency predictions. Our work provides a foundation for further research into saliency models that can deal with dynamic adaptations in interfaces, which are increasingly common, in particular on head-worn devices such as augmented and virtual reality interfaces~\cite{auitbelo22, lindlbauer_context_aware}. The \dataset{} dataset can also serve future research to investigate the impact of CL and HT on other aspects of eye gaze, such as saccadic behavior and can be useful for designing adaptive or dynamic interfaces that can direct users' attention to relevant information~\cite{feit20,muller22}

\subsection{Ethical Considerations, Limitations, and Future Work}
\changed{While dynamic adaptations such as visual highlighting offer significant potential to enhance user engagement and usability within interfaces, they can also be used to manipulate users attention and behavior. Therefore, we do not allow commercial use of our data and call on the research community to promote  the ethical use of gaze datasets and gaze research. On the other hand, our findings, and more generally predictive models of visual attention, could serve to help designers employ such features more responsibly and assess the impact on user attention already at design time to maintain their users' autonomy and trust. Similarly, they could serve to automatically detect deceptive or manipulative design practices on webpages as part of independent auditing processes or to display warnings to users. }

\changed{
Researchers utilizing the dataset must be mindful of potential biases in the collected gaze data, in particular given the comparatively small number of participants. While our participant pool is diverse in terms of gender and nationality, there is little variety in their age and education level. 
We encourage future work to collect larger datasets with more users and more variety of UIs, considering also 3D interfaces, or other cognitive aspects that might affect bottom-up attention. While these are costly to obtain for eye gaze, mouse input has been shown as a viable proxy~\cite{jiang2015salicon} for which our dataset could serve as a validation point. Following the data minimization principle, we did not collect any other participant-related information and only release anonymized gaze data and no demographic information to protect our participant's privacy. 
The study was approved by the ethical review board of the authors' university.}
% Our study was approved by ethical review board of the Faculty of Mathematics and Computer Science, Saarland University.

%While this number of participants should be increased by future work, the \dataset{} dataset provided a first indicator for the superior performance of saliency models that explicitly considered highlighting techniques and cognitive load of viewers. It could also serve future research to investigate the impact of CL and HT on other aspects of eye gaze, such as saccadic behavior and can be useful for designing adaptive or dynamic interfaces that can direct users' attention to relevant information~\cite{feit20,muller22}.

\subsection{The \dataset{} Dataset}
We publish a novel gaze dataset based on 150 unique webpages which were presented either in their original design, with a region permanently highlighted or the highlight appearing dynamically. Each stimulus was looked at under three cognitive load conditions (\textsf{Absent}, \textsf{Low}, and \textsf{High}). As a result, the \dataset{} dataset  consists of gaze coordinates, fixations, and saliency maps for 1350 unique stimuli: 450 stimuli viewed in three cognitive load conditions each, and correspondingly 450 viewed in three highlighting conditions, with observations from nine participants each.
%briefly summarize the study and our new dataset. Use this section to discuss limitations of the data collection and opportunities for future work to collect more extensive data and data on other congitive factirs?
%1.) EMPHASIZE novel dataset -- address existing gaps - who else (HCI, pysch, ET) can use these types of datasets -- further analysis - saccadic etc.
The \dataset{} dataset is available for non-commercial use at \url{https://osf.io/x8p9b/}, along with the accompanying code.
%\url{https://osf.io/x8p9b/?view_only=1dee01e9ce8442cfb7ef867a4bea290a})}. 

  %  5.) Limits \& Conclusion  --> dataset size \& sample; lab - real worl more attentional tunelling. dynamic highlight difficult to analyse with static comparatively due to uneven time of presentation, 
    
  %  3.) RQ2 How do the dynamics of highlighting and cognitive load impact the effectiveness of visual cues? Can we move some parts out of 4.3.1 \& 4.3.3 to here. IMPLICATIONS THAT dynamic copes well with CL --> discuss possible ethical issues vs usefulness. 4.3.2 to comment on comparing desgining for noticeability paper.
%    attentional tunneling: inability to travel larger distances away from the focus of attention under High Load.

%
 %   2.) RQ1: Move some parts of findings to : How does visual highlighting and cognitive load affect users’ viewing behavior of webpages: Unusual Results: why meanfixdur not significant for highlight? microsaccadic movement as they read? highlight is not always caught, so when it is not noticed people continue normal behaviour -- > no great impact on fixation behavior esp duration, so at the end compensatpry so no significant change in mean fixation duration.

% For Articles V8etra221-V8etra240 use:
\received{November 2023}
\received[revised]{January 2024}
\received[accepted]{March 2024}

\begin{acks}
This work is funded by DFG grant 389792660 as part of TRR~248 -- CPEC, see \url{https://perspicuous-computing.science}
\end{acks}

\bibliographystyle{ACM-Reference-Format}
% \bibliography{reference}


\end{document}

\title{Shifting Focus with HCEye: Supplementary Materials}

%\email{webmaster@marysville-ohio.com}
%\affiliation{%
%  \institution{Institute for Clarity in Documentation}
%  \streetaddress{P.O. Box 1212}
%  \country{USA}

\author{Anwesha Das}
\email{adas@cs.uni-saarland.de}
\orcid{0009-0006-8308-9961}
\authornote{These authors contributed equally to this work.}
\affiliation{%
 \institution{Saarland University, Saarland Informatics Campus}
 \city{Saarbrücken}
 \state{Saarland}
 \country{Germany}
 \postcode{66123}
}
\author{Zekun Wu}
\email{wuzekun@cs.uni-saarland.de}
\authornotemark[1]
\orcid{0000-0002-5233-2352}
\affiliation{%
 \institution{Saarland University, Saarland Informatics Campus}
 \city{Saarbrücken}
 \state{Saarland}
 \country{Germany}
 \postcode{66123}
}
\author{Iza Škrjanec}
\email{skrjanec@lst.uni-saarland.de}
\orcid{0009-0001-7044-8957}
\affiliation{%
 \institution{Saarland University, Language Science and Technology}
 \city{Saarbrücken}
 \state{Saarland}
 \country{Germany}
 \postcode{66123}
}
\author{Anna Maria Feit}
\email{feit@cs.uni-saarland.de}
\orcid{0000-0003-4168-6099}
\affiliation{%
 \institution{Saarland University, Saarland Informatics Campus}
 \city{Saarbrücken}
 \state{Saarland}
 \country{Germany}
 \postcode{66123}
}
\renewcommand{\shortauthors}{Anwesha Das et al.}
\maketitle

\section{User Study}

\subsection{Stimulus Design}

Our final dataset consists of 150 unique images:
\begin{itemize}
    \item From FiWI dataset \cite{shen2014webpage}:
    \begin{itemize}
        \item Number of images: 89
        \item Mean Feature Congestion Score (FC Score): $5.95$ ($SD = 0.99$, $Max = 9.25$)
    \end{itemize}
    
    \item From UEyes dataset \cite{jiang2023ueyes}:
    \begin{itemize}
        \item Number of images: 61
        \item Mean Feature Congestion Score (FC Score): $6.55$ ($SD = 1.10$, $Max = 9.61$)
    \end{itemize}
\end{itemize}

\begin{figure}[ht]
    \centering
    \begin{subfigure}[b]{0.49\textwidth}
        \centering
        \includegraphics[width=\textwidth]{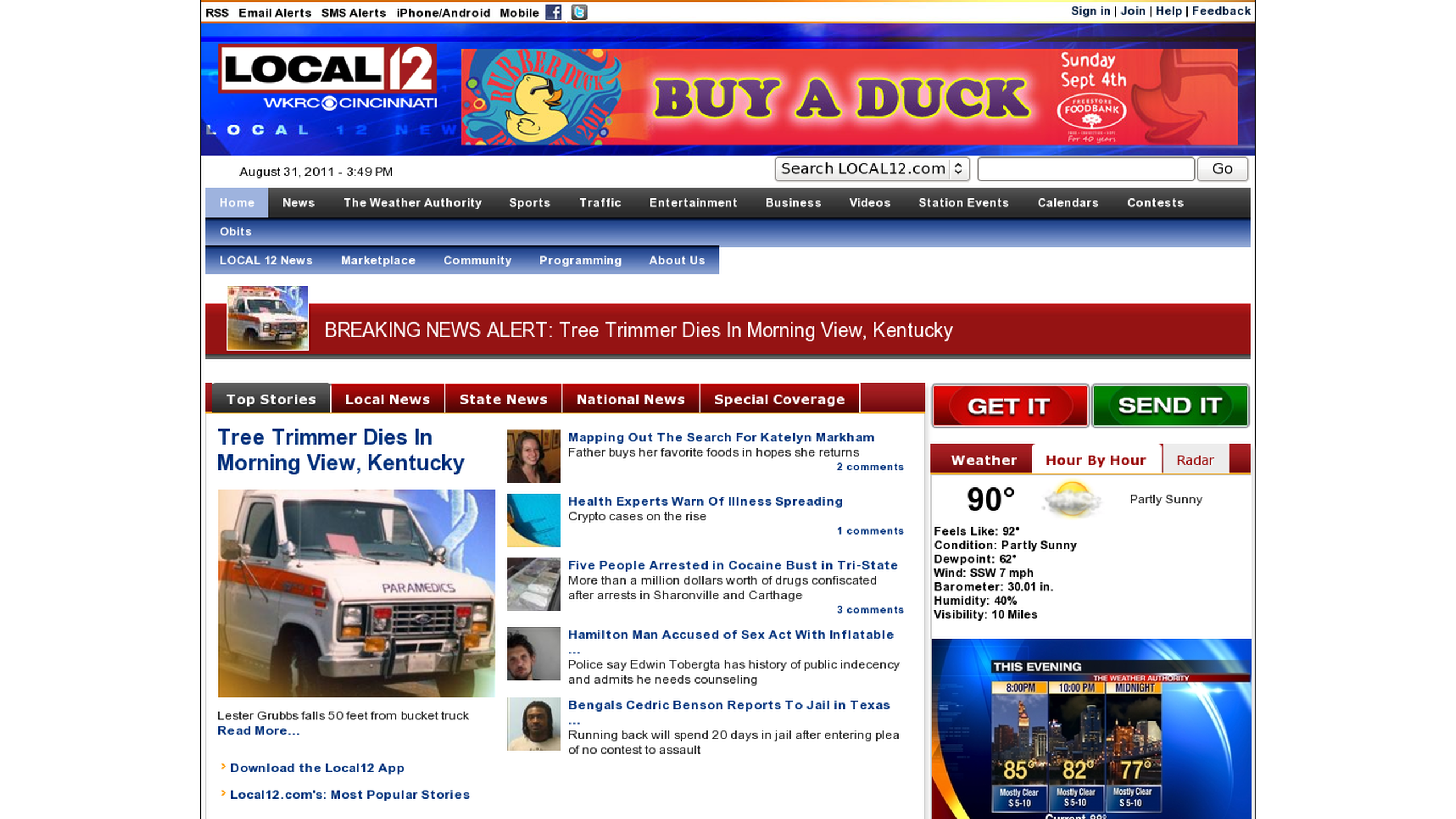}
        \caption{A complex visually cluttered UI with high FC Score.}
    \label{fig:HT_Stimulus_FC_subfig_a}
    \end{subfigure}
    \hfill
    \begin{subfigure}[b]{0.49\textwidth}
        \centering
        \includegraphics[width=\textwidth]{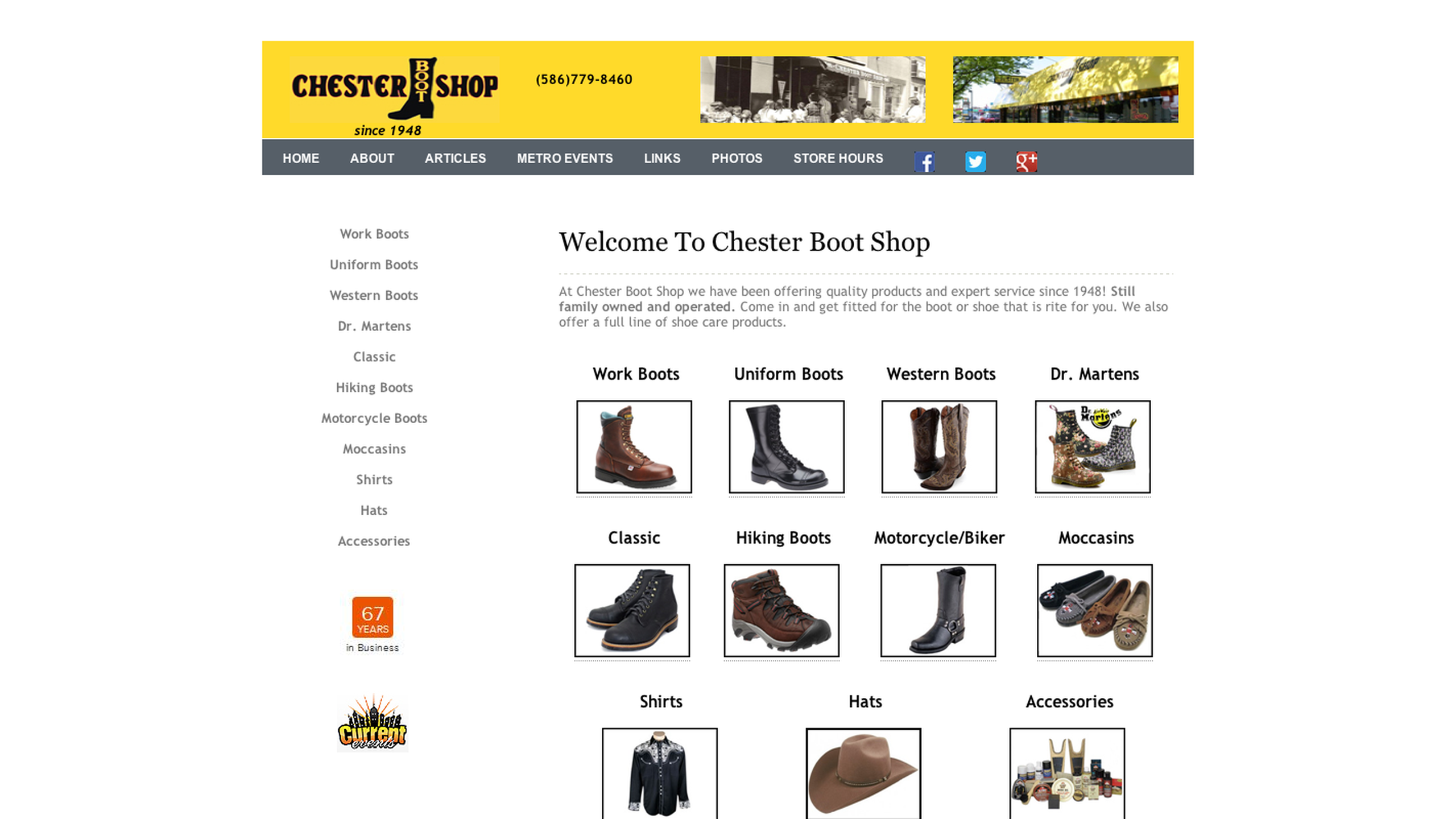}
        \caption{An UI with a simpler design UI and low FC Score.}
    \label{fig:HT_Stimulus_FC_subfig_b}
    \end{subfigure}
    \caption{Examples of two UI designs illustrating the use of FC Score \cite{rosenholtz2005feature} to identify complex versus simpler designs.}
    \label{fig:HT_Stimulus_FC}
\end{figure}

We used the Feature Congestion Score (FC Score) \cite{rosenholtz2005feature}, a measure of visual clutter, to select stimuli with complex designs containing multiple salient elements, interspersed with blocks of text and buttons featuring striking colors and images. \citet{oulasvirta2018aalto} provide an implementation and list thresholds (see \autoref{tab:feature_congestion}) of what constitutes \textit{complex} or poor visually cluttered UI designs as illustrated in Figures \ref{fig:HT_Stimulus_FC_subfig_a} and \ref{fig:HT_Stimulus_FC_subfig_b}.

\begin{table}[h]
\centering
\caption{Judgment and Score Ranges \cite{oulasvirta2018aalto} for Feature Congestion.}
\label{tab:feature_congestion}
\begin{tabular}{|c|c|}
\hline
\textbf{UI Design Judgment} & \textbf{Score Range} \\
\hline
Good & 0 to 3.771 \\
Fair & 3.772 to 5.5113 \\
Poor & \textbf{5.5114 and above} \\
\hline
\end{tabular}
\end{table}

\begin{figure}[ht]
    \centering
    \begin{subfigure}[b]{0.46\textwidth}
    \centering
    \includegraphics[width=\textwidth]{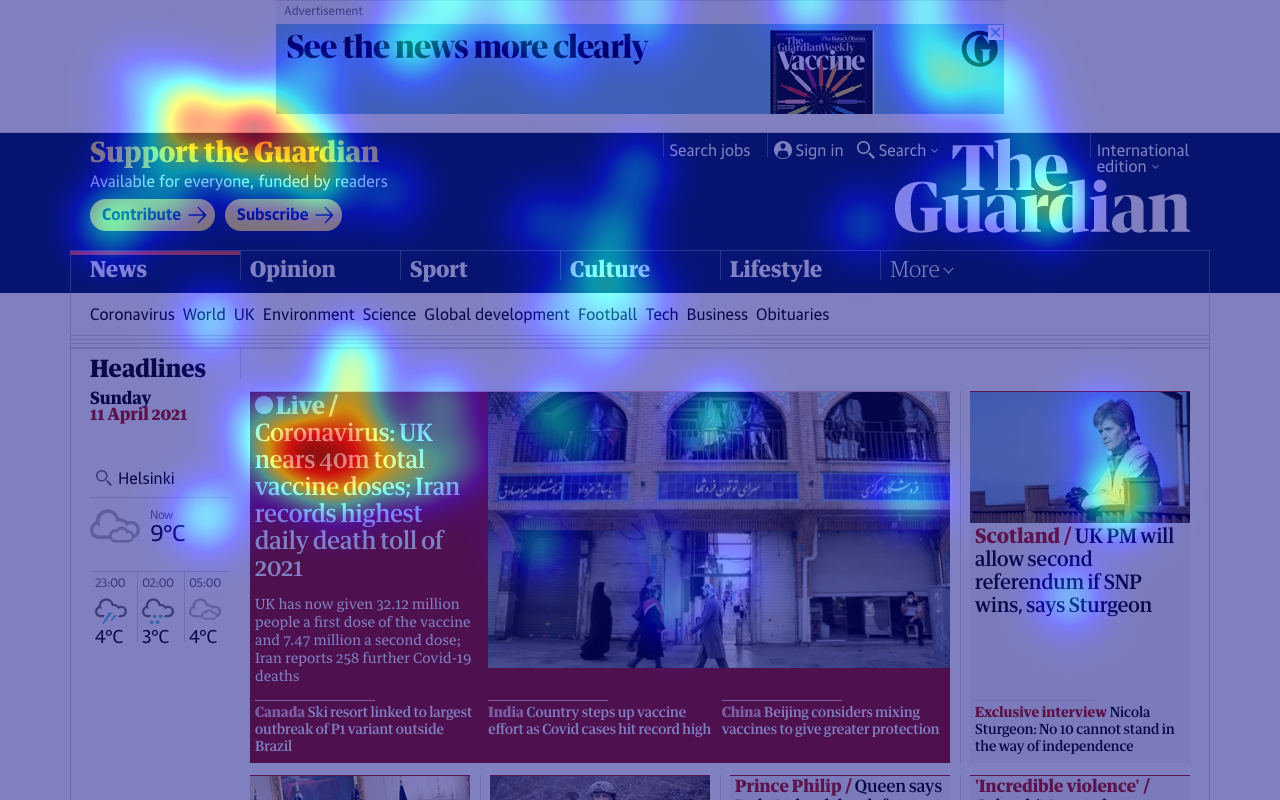}
        \caption{Heatmap of participants' gaze from UEyes.}
    \label{fig:2a}
    \end{subfigure}
    \hfill
    \begin{subfigure}[b]{0.51\textwidth}
        \centering
        \includegraphics[width=\textwidth]{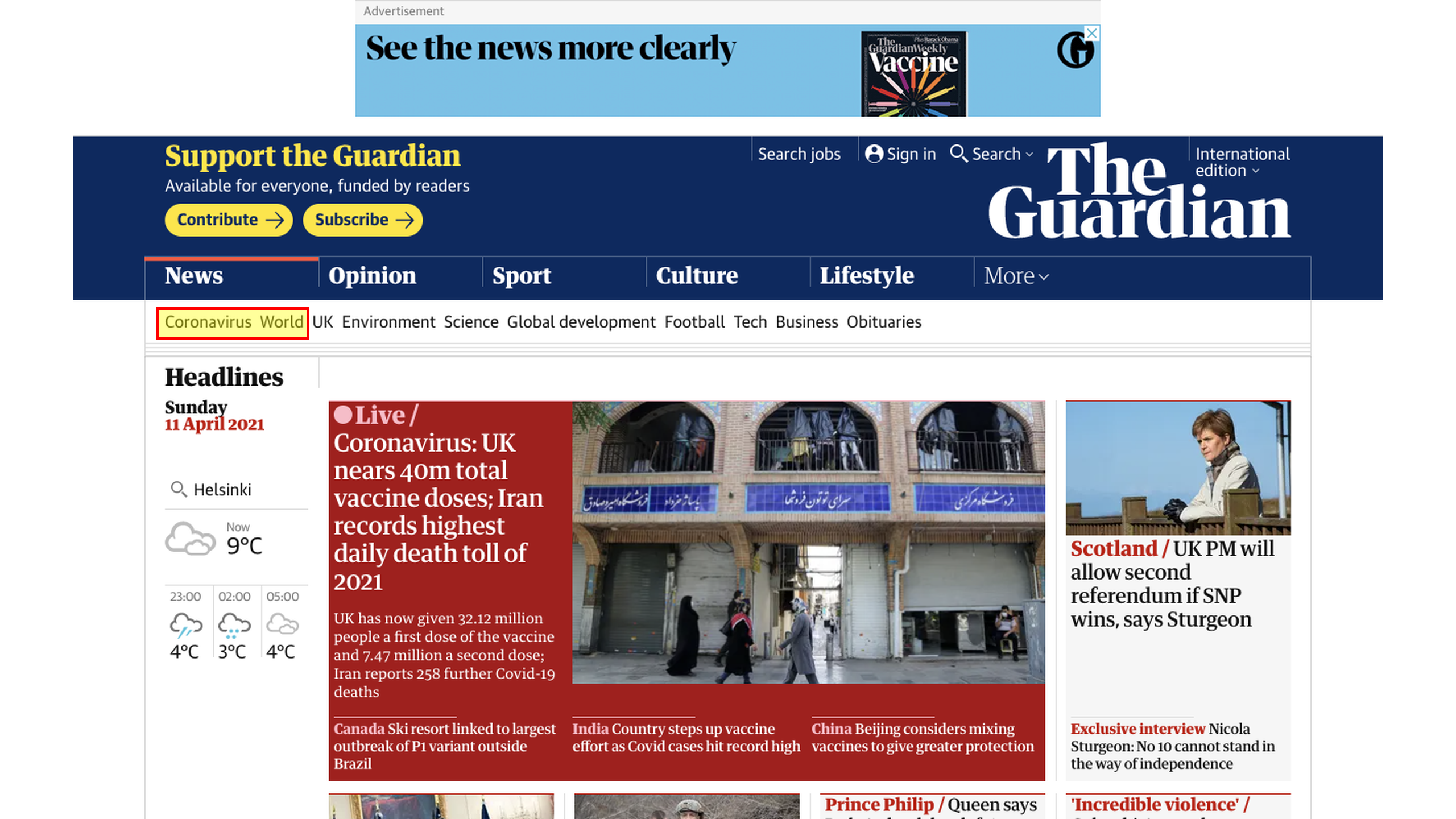}
        \caption{Highlight added to a previously inconspicuous region.}
    \label{fig:2b}
    \end{subfigure}
    \caption{Images with added yellow highlights to less-viewed regions, aiming to test efficacy of Highlighting Techniques \textbf{HT}.}
    \label{fig:HT_Applying}
\end{figure}

Each of these images are accompanied by two additional variants with a yellow highlight box -- the images with \textsf{Static} HT and \textsf{Dynamic} HT, resulting in a total of 450 stimuli images. The location of the highlights were spread through out various locations of the webpage stimuli. We visually inspected the heatmaps generated from or already provided by FiWI and UEyes datasets and added the visual highlight to a region where participants had not or very infrequently looked at, see \autoref{fig:HT_Applying}. This allowed us to artificially manipulate the saliency of regions that inherently would not attract users' gaze and thus would be good candidates to test the effectiveness of Highlighting Techniques \textbf{HT}.

% \begin{figure}[t]
%   \centering
% \includegraphics[width=1\textwidth]{figures/HT_Applying_2.png}
%   \caption{Left: Heatmap of participants' gaze from UEyes Right: Highlight added to a previously inconspicuous region}
%   \label{fig:HT_Applying}
% \end{figure}
% \newpage

\subsection{Study and Task Design}
\begin{figure}[t]
  \centering
\includegraphics[width=0.7\textwidth]{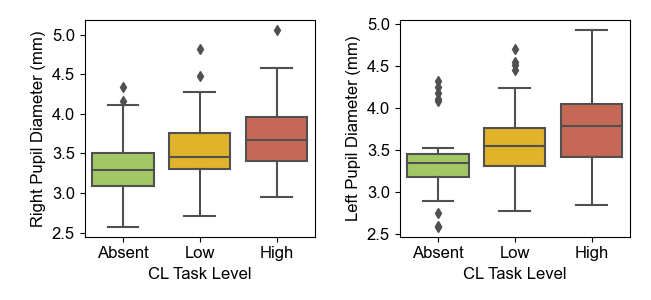}
  \caption{\textit{Left}: Increasing Right Pupil Diameter (mm) with increasing Task Difficulty; \space Right: Left Pupil Diameter (mm).}
\label{fig:CL_Pupil_Diameter}
\end{figure}

\changed{The decision to administer a mental arithmetic (counting) task was inspired by existing literature demonstrating its efficacy in inducing cognitive load, confirmed by various pupil-based metrics such as IPA, LHIPA, etc \cite{duchowski_index_pupillary, lindlbauer_context_aware, duchowski2020low}. The selection of the number 7 to count down by for the \textsf{High} CL task was based on preliminary pilot studies. Extensive research \cite{van2018pupil, beatty1966pupil} has consistently shown pupil dilation to serve as a reliable index of cognitive load, a finding corroborated also by our study. As illustrated in \autoref{fig:CL_Pupil_Diameter}, mean pupil diameter exhibits an upward trend with increasing task levels, indicating heightened cognitive load due to greater task complexity. Repeated Measures Analyses of Variance (ANOVAs) confirms significant variations in pupil diameter across task levels -- for the right pupil, $F(2, 78) = 91.682, p < 0.001$, and for the left pupil, $F(2, 78) = 84.863, p < 0.001$, supporting the observed trends depicted in \autoref{fig:CL_Pupil_Diameter}.}

\begin{figure}[t]
  \centering
\includegraphics[width=1\textwidth]{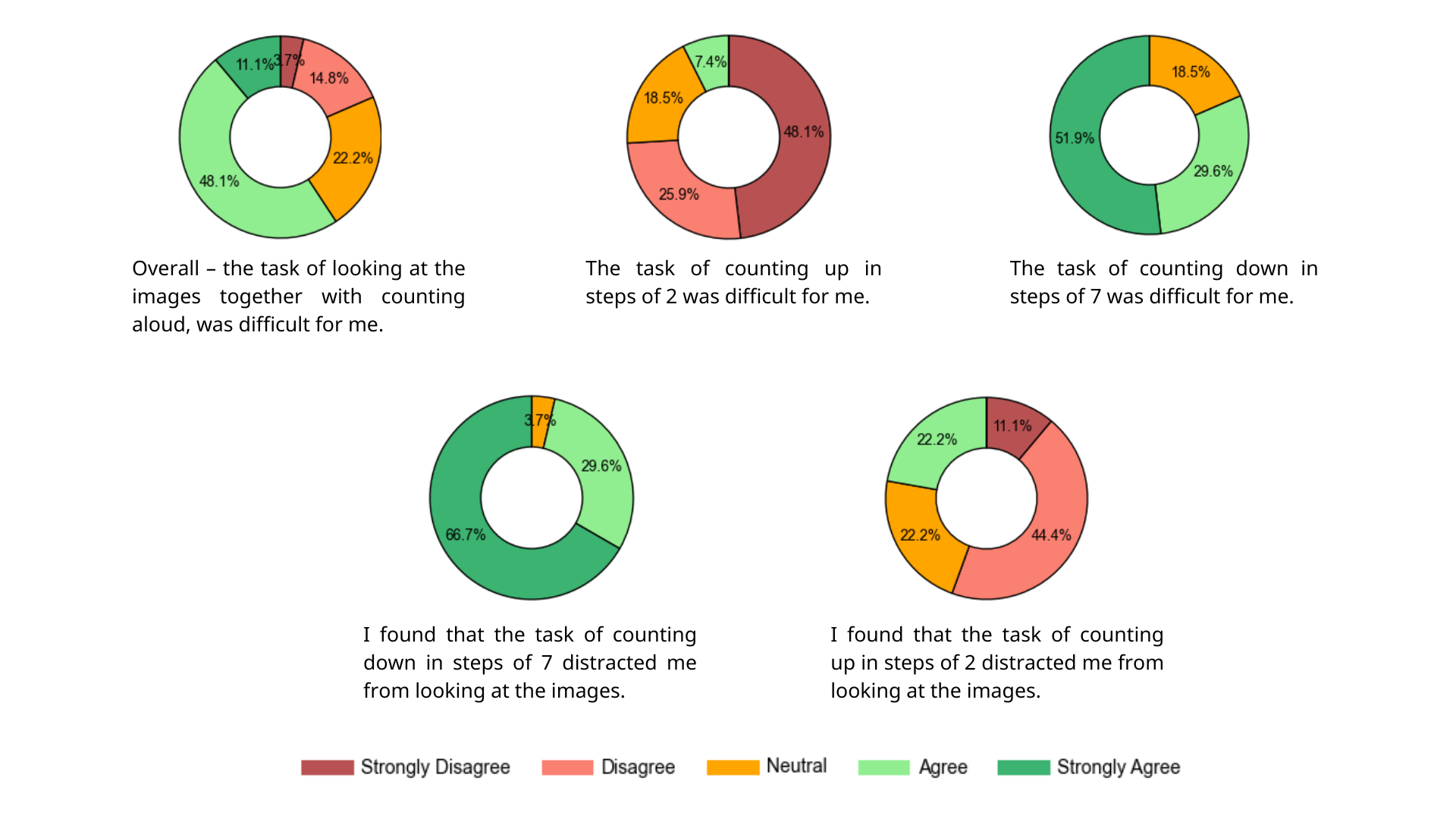}
  \caption{Participant Responses in Post-Study Questionnaire.}
\label{fig:Post Study Q}
\end{figure}

\changed{Responses gathered in the post-task questionnaire (shown in Figure \ref{fig:Post Study Q}) also confirm that participants found it challenging to simultaneously perform both the primary task (i.e., viewing stimuli `naturally') and the secondary task (i.e., verbalizing mental arithmetic calculations). 81.5\% of participants reported finding the countdown task in increments of 7 (\textsf{High} CL task) difficult, with 96.3\% indicating that it distracted their attention from the primary task. On the other hand, the task of counting upwards in intervals of 2 (\textsf{Low} CL task) was perceived as much easier, with only 7.4\% of participants reporting difficulty and 22.2\% noting distraction from the primary task.}

\newpage
\section{Shifting Attention}

Table \ref{tab:rank_table} tells us that the HR was engaged with more intently than the SR. $N_H$ increases with HT, and participants consistently dwell longer on the HR, reflected in the better average rank ($\mu_H$). $N_S$ decreases and so does its' average rank ($\mu_S$), with the presence of a HT. Naturally, in \textsf{Absent} HT condition, $N_S$ is at its highest, and $N_H$ lowest but \textsf{Dynamic} or \textsf{Static} HT compete for participants' attention  resulting in the shift from SR to HR. The higher values of $\mu_H$ (>$\mu_S$)
also indicate that information processing in the HR is more thorough, with HT having engaged viewers top-down attention. 

\begin{figure}[h]
  \centering
  \includegraphics[width=0.5\linewidth]{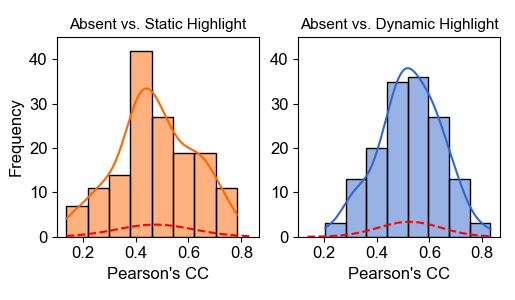}
  \caption{Pearson's CC for \textsf{Absent} vs. \textsf{Static} and \textsf{Absent} vs. \textsf{Dynamic} HT conditions.}
  \label{fig:pearson_CC_values}
\end{figure}

\begin{figure}[ht]
    \centering
    \begin{subfigure}[b]{0.325\textwidth}
        \centering
        \includegraphics[width=\textwidth]{figures/Amazon(1).png}
        \caption{\textsf{Absent} HT}
    \end{subfigure}
    \hfill
    \begin{subfigure}[b]{0.325\textwidth}
        \centering
        \includegraphics[width=\textwidth]{figures/Amazon(1)_highlight.png}
        \caption{\textsf{Static} HT}
    \end{subfigure}
    \hfill
    \begin{subfigure}[b]{0.325\textwidth}
        \centering
        \includegraphics[width=\textwidth]{figures/Amazon(1)_dynamic.png}
        \caption{\textsf{Dynamic} HT}
    \end{subfigure}
    \medskip
    % Image Cards 2
    \begin{subfigure}[b]{0.325\textwidth}
        \centering
        \includegraphics[width=\textwidth]{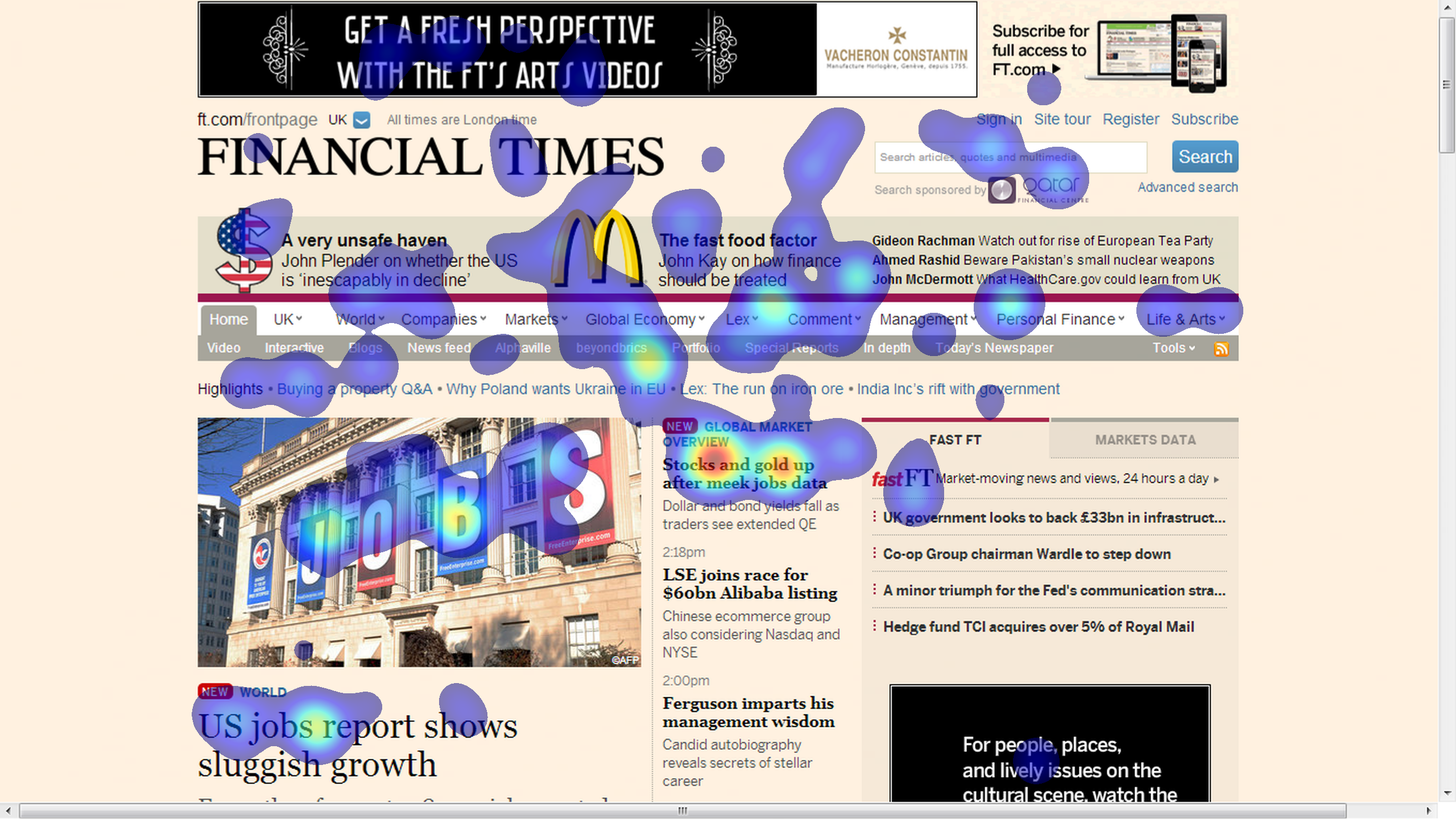}
        % \caption{Financial Times}
    \end{subfigure}
    \hfill
    \begin{subfigure}[b]{0.325\textwidth}
        \centering
        \includegraphics[width=\textwidth]{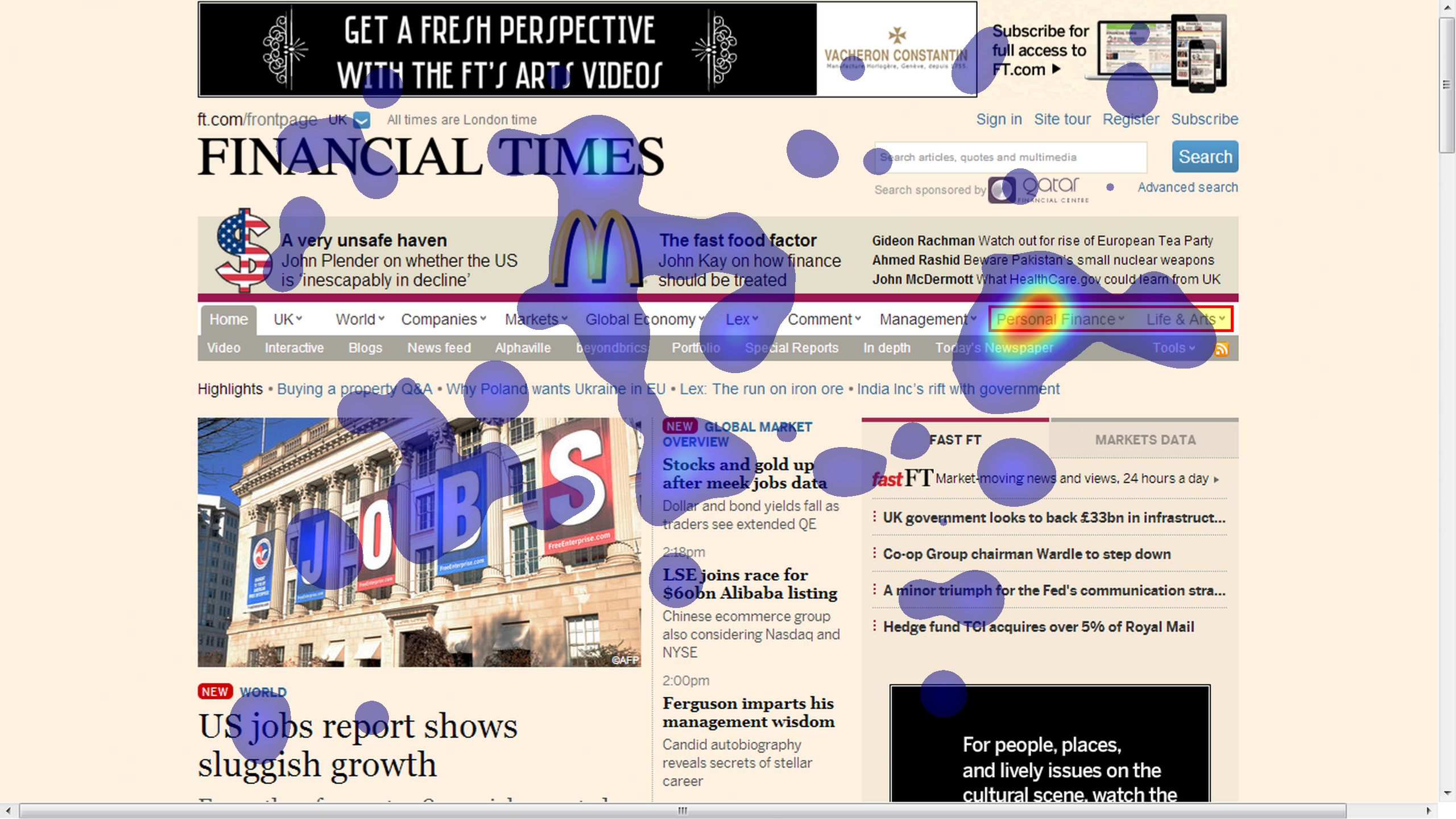}
        % \caption{Financial Times highlight}
    \end{subfigure}
    \hfill
    \begin{subfigure}[b]{0.325\textwidth}
        \centering
        \includegraphics[width=\textwidth]{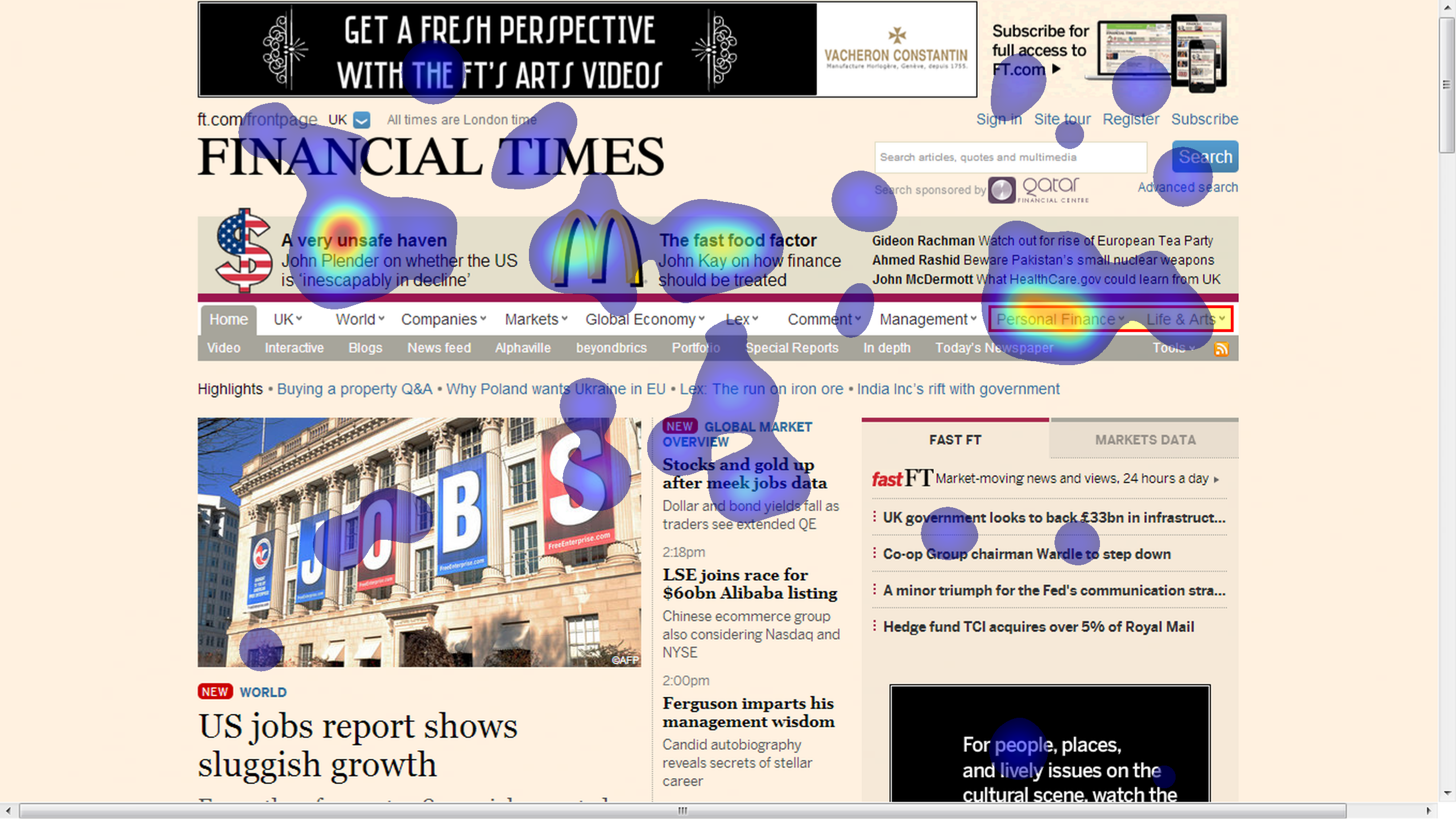}
        % \caption{Financial Times dynamic}
    \end{subfigure}
    \caption{Aggregated Heatmaps of participants viewing two different webpage stimuli under various HT conditions, showcasing how viewing behaviour changes with the presence of visual highlights.}
    \label{fig:QualitativeShiftingAttention}
\end{figure}

This shift of attention explains the moderately low CC values when we compare the distributions of gaze behaviour of participants in \textsf{Absent} vs \textsf{Static} HT. We see, in Figure \ref{fig:pearson_CC_values}, that the distribution for the \textsf{Dynamic} HT achieves stronger correlation to \textsf{Absent} HT due to the latency of the arrival of the HT (3s later) allowing participants to exhibit gaze behaviour similar to how they would without any HT present. Paired-t-tests confirm this increase in CC with \textsf{Dynamic} HT being significantly different $(t =- 4.3988, p<0.00001)$ from \textsf{Static} HT. This emphasizes the need for further separate saliency modelling for UIs with dynamic interactions, which the changing gaze behaviour exhibited by our qualitative images of participants' heatmaps under each HT condition also supports (see Figure \ref{fig:QualitativeShiftingAttention}).

\begin{table}[h]
  \centering
  \begin{minipage}{1\textwidth}  % Adjust width as needed
    \fontsize{9}{10}\selectfont
    \sffamily
    \begin{tabular}{|p{1.5cm}|p{3.5cm}|p{3.5cm}|p{3.5cm}|}  % Adjust column widths as needed
      \hline
       & Absent Highlight & Static Highlight & Dynamic Highlight \\
      \hline
      Absent CL & \begin{tabular}{@{}p{2cm}@{}p{2cm}@{}} $N_{S} = 78.67$ & $\mu_{S} = \underline{2.051}$ \\ $N_{H} = 9.33$ & $\mu_{H} = 3.0$ \\ \end{tabular} & \begin{tabular}{@{}p{2cm}@{}p{2cm}@{}} $N_{S} = 72.67$ & $\mu_{S} = \underline{2.110}$ \\ $N_{H} = 80.67$ & $\mu_{H} = 2.116$ \\ \end{tabular} & \begin{tabular}{@{}p{2cm}@{}p{2cm}@{}} $N_{S} = 59.33$ & $\mu_{S} = 2.978$ \\ $N_{H} = 92.67$ & $\mu_{H} = \underline{1.813}$ \\ \end{tabular} \\
      \hline
      Low CL & \begin{tabular}{@{}p{2cm}@{}p{2cm}@{}} $N_{S} = 79.33$ & $\mu_{S} = \underline{1.748}$ \\ $N_{H} = 10.67$ & $\mu_{H} = 3.0$ \\ \end{tabular} & \begin{tabular}{@{}p{2cm}@{}p{2cm}@{}} $N_{S} = 69.33$ & $\mu_{S} = 2.01$ \\ $N_{H} = 86.67$ & $\mu_{H} = \underline{1.646}$ \\ \end{tabular} & \begin{tabular}{@{}p{2cm}@{}p{2cm}@{}} $N_{S} = 58.0$ & $\mu_{S} = 2.414$ \\ $N_{H} = 79.33$ & $\mu_{H} = \underline{1.706}$ \\ \end{tabular} \\
      \hline
      High CL & \begin{tabular}{@{}p{2cm}@{}p{2cm}@{}} $N_{S} = 76.0$ & $\mu_{S} = \underline{1.43}$ \\ $N_{H} = 5.33$ & $\mu_{H} = 2.125$ \\ \end{tabular} & \begin{tabular}{@{}p{2cm}@{}p{2cm}@{}} $N_{S} = 65.33$ & $\mu_{S} = 1.786$ \\ $N_{H} = 69.33$ & $\mu_{H} = \underline{1.683}$ \\ \end{tabular} & \begin{tabular}{@{}p{2cm}@{}p{2cm}@{}} $N_{S} = 50.0$ & $\mu_{S} = 2.253$ \\ $N_{H} = 83.33$ & $\mu_{H} = \underline{1.808}$ \\ \end{tabular} \\
      \hline
    \end{tabular}
    \caption{\textit{Shift in Attention}: $N_{S}$ \& $N_{H}$ are the percentage of times that the SR \& HR respectively were attended to compared to all the regions that were looked at by participants; $\mu_{S}$ \& $\mu_{H}$ are the average ranks of SR \& HR (1 being the best) with the better rank ($\mu_{S}$ compared to $\mu_{H}$) \underline{underlined} in the Table above.}
    \label{tab:rank_table}
  \end{minipage}
\end{table}

\section{Model Input Structure for Dynamic Highlight}
The original SimpleNet that employed the PNAS network is designed to process standard 3-channel RGB images. In our work, we propose to adjust it to accommodate the two stacked image inputs from the \textsf{Dynamic} highlight condition. Specifically, as indicated in the Figure \ref{fig:arch7} below, we needed to process pair input and then forward them into the PNAS encoder as a unified model framework. This requirement led us to add the Channel Reduction layer, which effectively compresses the 6-channel input (resulting from concatenating two 3-channel images) into a 3-channel format, and using Batch Normalization, which normalizes the output of the channel reduction layer, stabilizing the learning process and potentially improving the training efficiency. 

\begin{figure}[ht]
\setlength{\intextsep}{2pt} % Adjust this value as needed
\centering
\includegraphics[width=0.6\linewidth]{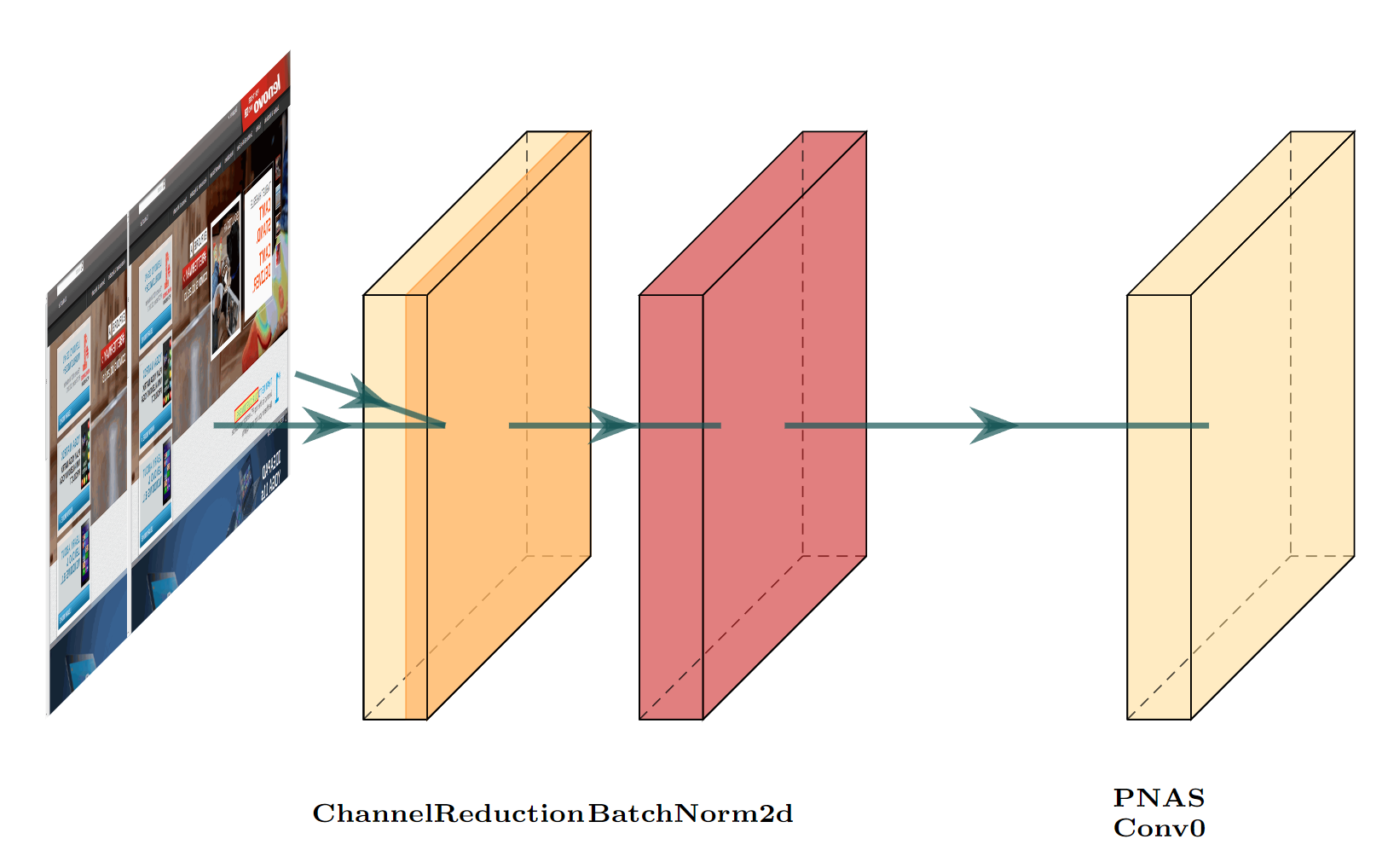}
\caption{Channel Reduction and Batch Normalization for Pair Input from the \textsf{Dynamic} Highlight.}
\label{fig:arch7}
\end{figure}

\section{Saliency Predication Qualitative Results}
\begin{figure}[ht]
\setlength{\intextsep}{2pt} % Adjust this value as needed
\centering
\begin{multicols}{3}
    \begin{minipage}{\linewidth}
        \centering
        \includegraphics[width=\linewidth]{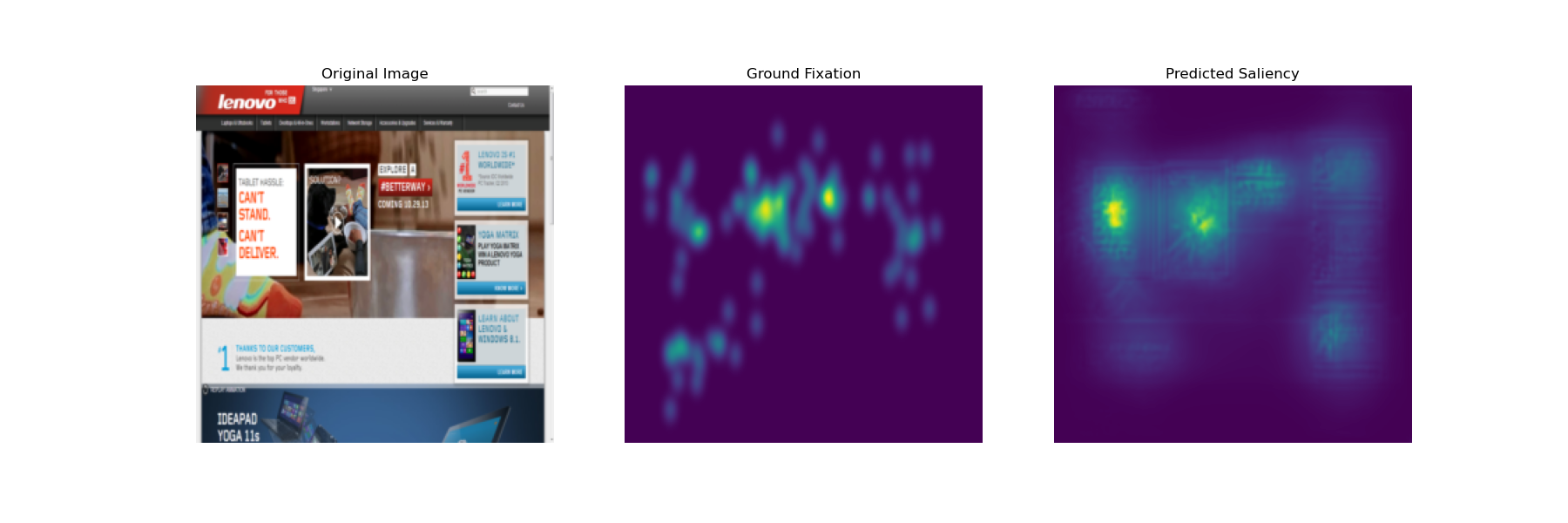}
        \subcaption*{With Highlight \textsf{Absent}}
    \end{minipage}%

    \begin{minipage}{\linewidth}
        \centering
        \includegraphics[width=\linewidth]{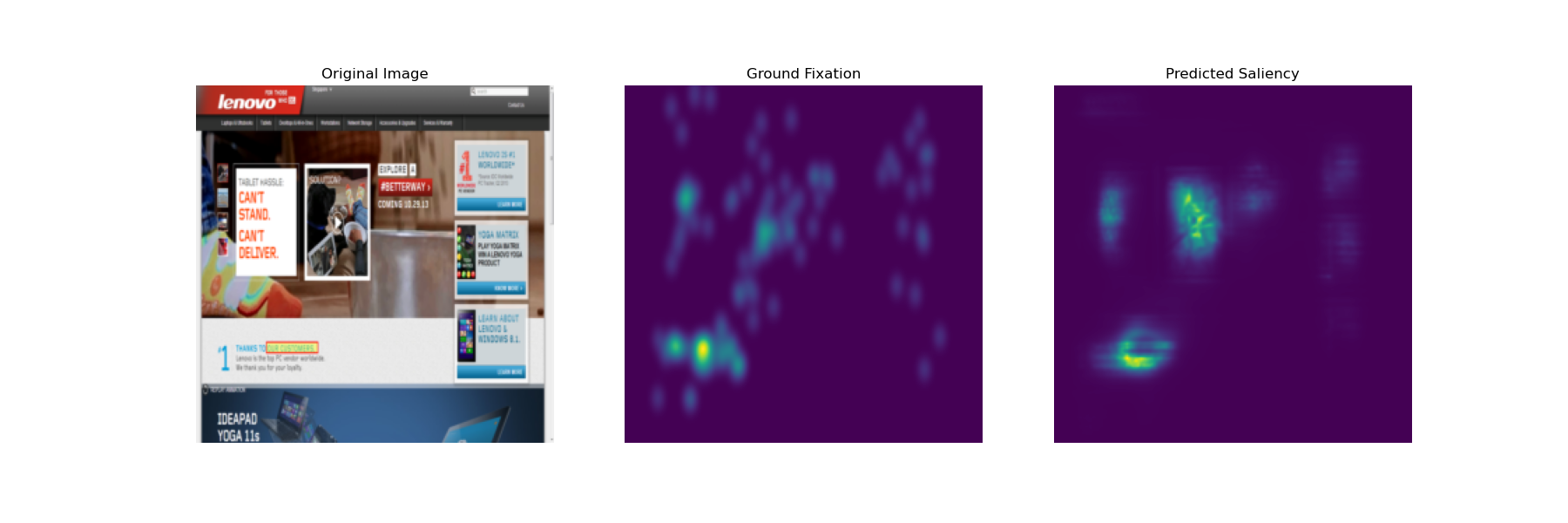}
        \subcaption*{With \textsf{Static} Highlight}
    \end{minipage}%

    \begin{minipage}{\linewidth}
        \centering
        \includegraphics[width=\linewidth]{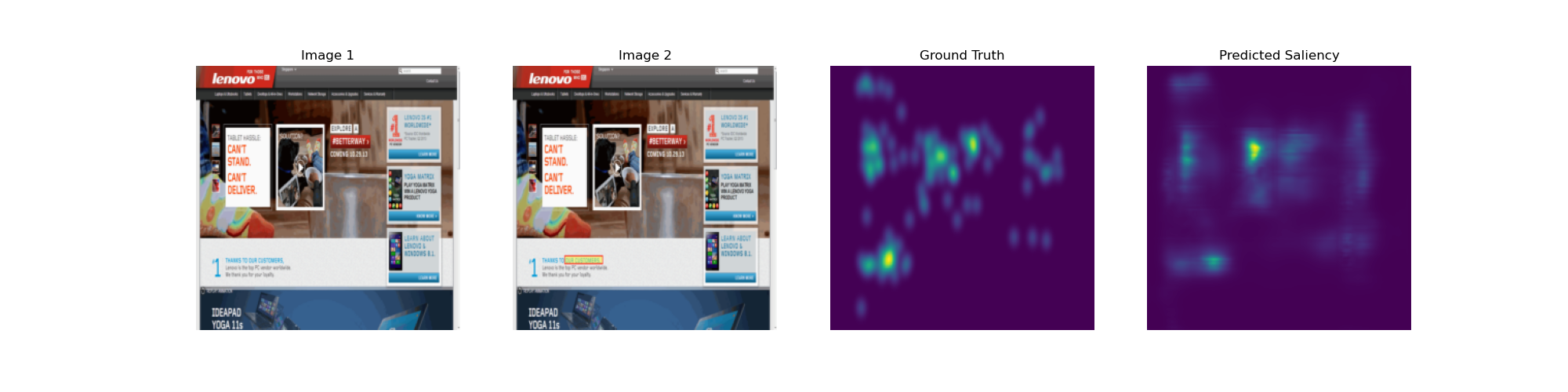}
        \subcaption*{With \textsf{Dynamic} Highlight}
    \end{minipage}%
\end{multicols}
\caption{Saliency prediction comparison across different highlight conditions (Fine-turned).}
\label{fig:highlight_conditions}
\end{figure}

\autoref{fig:highlight_conditions} presents a comparison of groundtruth fixation maps and our model's predictions under various highlighting conditions. The saliency maps generated by our model reveal distinct patterns of visual attention depending on the highlight condition:

Absent Highlight: In the \textsf{Absent} Highlight condition, the model's predicted saliency map is dispersed, indicating a general visual exploration across the UI without specific focal points. This dispersion suggests that in the absence of visual cues, attention is distributed more evenly across the UI.

Static Highlight: With the introduction of \textsf{Static} highlights, the model predicts a more concentrated saliency in these areas. This concentration demonstrates that static visual cues are effective in attracting viewer attention and directing it towards targeted regions on the UI.

Dynamic Highlight: The \textsf{Dynamic} highlight condition presents an interesting case. The model predicts intensified saliency around dynamic elements, while also maintaining a somewhat dispersed attention across the image, unlike in the static case. This pattern indicates that the model is highly responsive to changes and movement within the UI, recognizing these dynamic elements as potent attractors of visual attention.

\begin{figure}[ht]
\setlength{\intextsep}{2pt} % Adjust this value as needed
\centering
\begin{multicols}{3}
    \begin{minipage}{\linewidth}
        \centering
        \includegraphics[width=\linewidth]{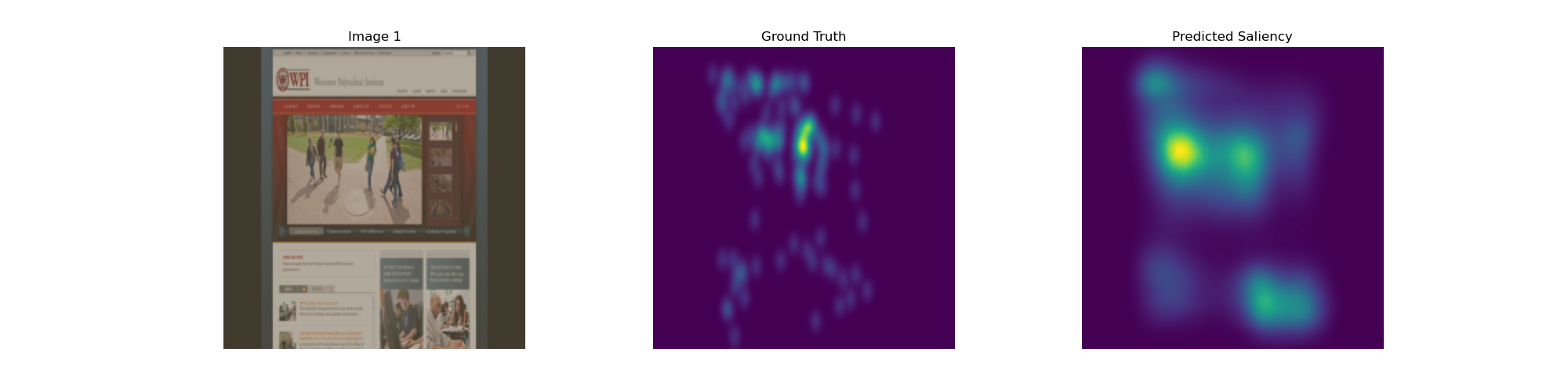}
        \subcaption*{With Highlight \textsf{Absent}}
    \end{minipage}%

    \begin{minipage}{\linewidth}
        \centering
        \includegraphics[width=\linewidth]{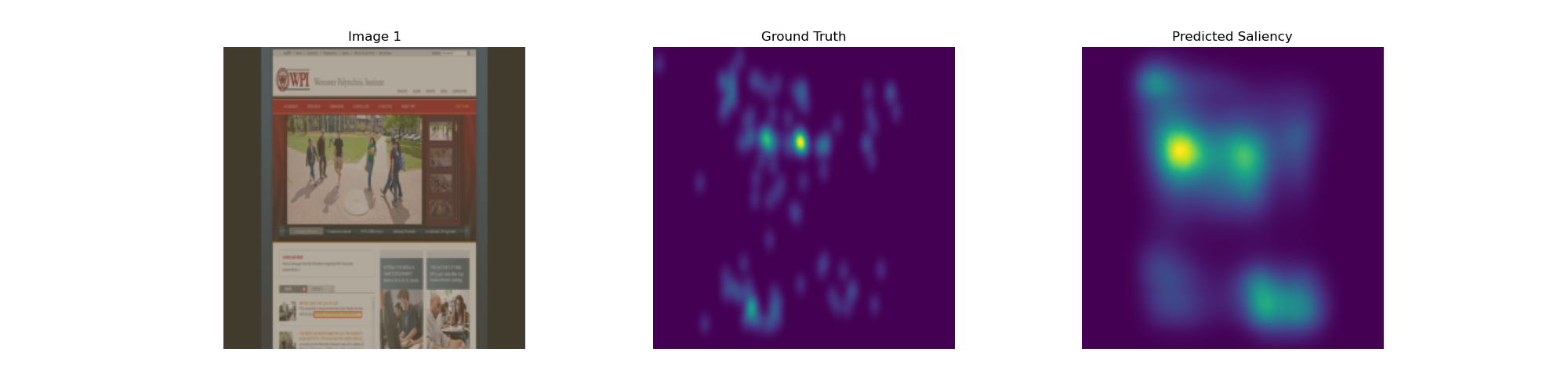}
        \subcaption*{With \textsf{Static} Highlight}
    \end{minipage}%

    \begin{minipage}{\linewidth}
        \centering
        \includegraphics[width=\linewidth]{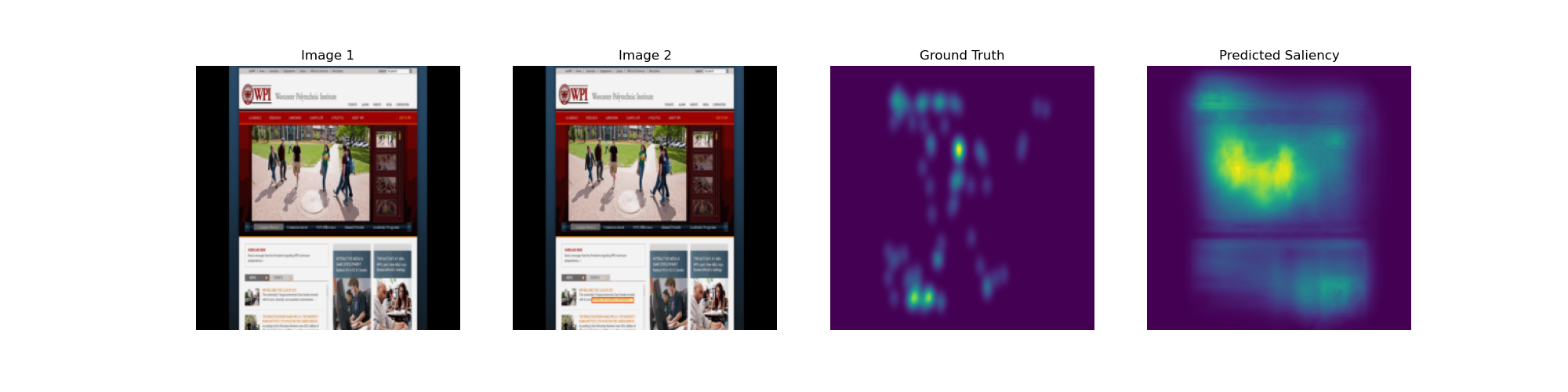}
        \subcaption*{With \textsf{Dynamic} Highlight}
    \end{minipage}%
\end{multicols}
\caption{Saliency prediction comparison across different highlight conditions (Pre-trained).}
\label{fig:highlight_conditions_org}
\end{figure}

In contrast, \autoref{fig:highlight_conditions_org} displays the predictions made by the pre-trained model under the same highlight conditions. Unlike the fine-tuned model, the pre-trained model predominantly focuses its predictive visual attention on human objects within the images, located in both the top and bottom right of the webpage. It notably ignores the highlighted text in the bottom left. This comparison underscores the effectiveness of training the SimpleNet model with visual stimuli and corresponding gaze data in various highlight conditions, demonstrating a significant shift in the model's attentional focus due to the fine-tuning process.

\bibliographystyle{ACM-Reference-Format}